\documentclass{aa}  

\usepackage{graphicx}
\usepackage{txfonts}
\usepackage[]{natbib}
\usepackage[]{subfig}
\usepackage{color}
\usepackage{amsmath}
\usepackage{ulem}
\definecolor{gris}{gray}{0.5}

\newcommand{\kms}{km~s$^{-1}$}
\begin{document}

 \title{The HH30 edge-on T Tauri star}

   \subtitle{A rotating and precessing monopolar outflow scrutinized by ALMA}

   \author{F. Louvet
          \inst{1}
          \and
          C. Dougados\inst{2,1,3}
       \and 
           S. Cabrit\inst{4,3}
                  \and
           D. Mardones\inst{1}
             \and
           F. M\'enard\inst{2,1,3}
    \and 
        B. Tabone\inst{4}
    \and 
         C. Pinte\inst{2,1,3,5}
       \and
                  W. R. F. Dent\inst{6}
        }

   \institute{Departamento de Astronomia de Chile, Universidad de Chile, Santiago, Chile
              \email{flouvet@das.uchile.cl}
          \and
             UMI-FCA, CNRS/INSU, France (UMI 3386)            
         \and
             Univ. Grenoble Alpes, CNRS, IPAG, 38000 Grenoble, France
        \and
             Laboratoire d’Etudes du Rayonnement et de la Matière en Astrophysique et Atmosphères (LERMA) - Observatoire de Paris-Meudon, France            
\and
        Monash Centre for Astrophysics (MoCA) and School of Physics and Astronomy, Monash University, Clayton Vic 3800, Australia
        \and
          ALMA/ESO, Alonso de Cordova 3107, Santiago, Chile
           }

   \date{Received September 15, 1996; accepted March 16, 1997}

 
  \abstract
   {The disk-outflow connection is thought to play a key role in extracting excess angular momentum from a forming protostar. HH30 is a rare and beautiful example of a pre-main sequence star exhibiting a flared edge-on disk, an optical jet, and a CO molecular outflow, making this object a case study for the disk-jet-outflow paradigm.}
   {We aim to clarify the origin of the small-scale molecular outflow of HH30 and its link and impact on the accretion disk.}
   {We present ALMA 0.25$^{\prime\prime}$ angular resolution observations of the circumstellar disk and outflow around the T Tauri star HH30 in the dust continuum at 1.33 mm and of the molecular line transitions of $^{12}$CO(2-1) and $^{13}$CO(2-1). We performed a disk subtraction from the $^{12}$CO emission, from which we analysed the outflow properties in detail in the altitudes z$\lesssim$250~au. We fit the transverse position-velocity diagrams across the $^{12}$CO outflow to derive the ring positions and projected velocity components (including rotation). We use the results of these fits to discuss the origin of the CO outflow.}
   {The 1.3~mm continuum emission shows a remarkable elongated morphology along PA=31.2$^{\circ}$~$\pm$~0.1$^{\circ}$ that has a constant brightness out to a radius of r=75~au. The emission is marginally resolved in the transverse direction, implying an intrinsic vertical width $\leq$~24~au and an inclination to the line-of-sight ${\rm i}~\ge~84.8^{\circ}$. The $^{13}$CO emission is compatible with emission from a disk in Keplerian rotation, in agreement with the previous findings. The monopolar outflow, detected in $^{12}$CO, arises from the north-eastern face of the disk from a disk radius r~$\le$~22~au and extends up to 5$^{\prime\prime}$ (or 700~au) above the disk plane. We derive a lower limit to the total mass of the CO cavity/outflow of $1.7\times10^{-5}$ M$_{\odot}$. The CO cavity morphology is that of a hollow cone with semi-opening angle $\sim$35$^\circ$. The derived kinematics are consistent with gas flowing along the conical surface with constant velocity of 9.3~$\pm$~0.7~\kms. We detect small rotation signatures (V$_\phi \sin{\rm i}\in[0.1;0.5]$~\kms) in the same sense as the underlying circumstellar disk. From these rotation signatures we infer an average specific angular momentum of the outflow of 38~$\pm$~15~au~\kms at altitudes z~$\le$~250~au. We also report the detection of small amplitude wiggling (1.2$^{\circ}$) of the CO axis around an average inclination to the line of sight of i=91$^{\circ}$.}
  {The derived morphology and kinematics of the CO cavity are compatible with expectations from a slow disk wind, originating either through photo-evaporation or magneto-centrifugal processes. Under the steady assumption, we derive launching radii in the range 0.5-7~au. In that scenario, we confirm the large minimum mass flux of 9$\times 10^{-8}$ M$_\odot$ yr$^{-1}$  for the CO wind. The wind would therefore extract a significant amount of the accreted mass flux through the disk and would likely play a crucial role in the disk evolution. If the CO flow originates from a steady-state disk wind, our ALMA observations rule out the 18~au binary orbital scenario previously proposed to account for the wiggling of the optical jet and favour instead a precession scenario in which the CO flow originates from a circumbinary disk around a close (separation $\leq$~3.5~au) binary. Alternatively, the CO outflow could also trace the walls of a stationary cavity created by the propagation of multiple bow shocks. Detailed numerical simulations are under way to fully test the entrainment hypothesis.}


   \maketitle
%

\section{Introduction}

A necessary prerequisite to understand the formation of stars is the comprehension of the complex processes linking the collapsing molecular core, the protostar, its circumstellar disk, and the bipolar jets and outflows that expel material. Together, these processes regulate the protostar fragmentation and the mass that the protostar(s) acquires; plus, they appear to be key for the existence and morphology of planetary systems. Among these processes, the initial amount, evolution, and re-distribution of angular momentum appear to be fundamental. Part of the excess of angular momentum may be carried away by jets and outflows and, thus, provide a solution to the angular momentum problem in star formation \citep[e.g.][]{ray07}. Yet, the exact link between jets/flows and the accretion disk is still a critical issue in contemporary physics. One attractive possibility is a transfer of angular momentum from the disk to the jets/outflows by means of magneto-centrifugal forces, such that circumstellar material may continue to accrete onto the central object \citep[e.g.][]{blandford82}. Exactly where and how this transfer occurs, and how it impacts the disk physics, is however still hotly debated \citep{ferreira06, pudritz07, shang07, romanova09, cabrit09}.  

Measurements of angular momentum have been reported for the jets in various evolutionary phases from Class 0 \citep{lee08} to Class I \citep{chrysostomou08}, and during the T Tauri phase \citep{bacciotti02,woitas05,coffey04,coffey07}. Under the steady mass loss assumption, these signatures imply a jet launching radius in the inner 0.1-3~au of the disk and suggest that magneto hydrodynamic (MHD) winds could fully drive the accretion in these regions. However, \cite{louvet16b} showed that for the T Tauri star Th28, the rotation sense of the disk is opposite to that of the transverse velocity shifts that were previously detected with the Hubble Space Telescope (HST) in the optical jet of this source \citep{coffey07}. That second example of counter-rotation together with RW Aur \citep{coffey04,cabrit06} suggests that the steady assumption may not hold and casts doubt on the ability to derive constraints on the launching radii of jets derived from optical rotation signatures.

Slower molecular outflows may also play an important role in reducing the angular momentum from the disk/protostar system.
The traditional interpretation of CO flows in terms of swept-up ambient matter has recently been challenged by the detection of small-scale, V-shaped CO cavities in evolved Class II sources. In these sources, no obvious envelope is present for entrainment and the cavity base originates from within the circumstellar disk \citep[e.g. in HH30][]{pety06}. Alternatively these small-scale cavities could trace disk winds generated either by magneto-hydrodynamical processes or by photo-evaporation of the outer disk atmosphere.
The MHD disk winds are efficient at extracting angular momentum, while photo-evaporative flows have potentially a strong influence on disk gas dissipation processes. Recent studies have reported tentative rotation signatures in low-velocity Class~0 and Class~I molecular outflows at a level consistent with MHD disk winds \citep{launhardt09,zapata09,bjerkeli16,tabone17,hirota17}. In all these embedded sources however, the entrainment scenario cannot be fully excluded.
In this article, we present a detailed study of the small-scale CO cavity/molecular outflow from the edge-on Class II source HH30 conducted with ALMA.

The Herbig-Haro (HH) object 30 \citep{mundt83} is a young solar-type star devoid of an envelope located in the dark molecular cloud L1551 at a distance of $\sim$140 pc \citep{kenyon94} in Taurus. The HH 30 exciting source is an optically invisible star \citep{vrba85} that is highly extinguished by an edge-on disk \citep{burrows96,stapelfeldt99}, which extends up to a radius of $\sim$250 au perpendicular to the jet and divides the surrounding reflection nebulosity into two lobes. Recent interferometric observations in $^{13}$CO (J = 2-1) are consistent with a gaseous disk in Keplerian rotation around an enclosed mass of 0.45 $\pm$ 0.04 M$_\odot$ that corresponds to a typical T Tauri star with spectral class M0 $\pm$ 1 \citep[][hereafter P06]{pety06}. HH30 is considered as a prototype disk/jet/outflow system. Its impressive bipolar jet has a total angular size of 7' \citep{anglada07}. The overall HH 30 jet structure can be well described by a wiggling ballistic jet, whose knots have velocities included between 100 km s$^{-1}$ and 300 km s$^{-1}$ \citep{estalella12}. \cite{anglada07} suggested that the wiggling arises either from the orbital motion of the jet source around a primary or from precession of the jet axis because of the tidal effects of a companion. In the first scenario, the companion would be orbiting at $\sim$18 au in a 53-year period, whereas in the second scenario the companion would be orbiting at less than 1 au  in less than a year. Interferometric imaging in the continuum at $\lambda$ = 1.3 mm resolved a region of reduced brightness at the centre of the system, suggesting that the disk of HH30 is truncated at an inner radius of 37$\pm$4 au \citep{guilloteau08}; this implies that the wiggling of the jet would be due to orbital motion. The molecular gas around HH30 was studied by \cite{pety06} with the Plateau de Bure interferometer (hereafter PdBI) at an angular resolution of $\sim$1.4$^{''}$. The P06 work showed that the disk of HH30 is in Keplerian rotation with its rotation vector pointing towards the north-eastern jet. Furthermore, P06 demonstrated that the outflow of HH30 is expending in the plane of the sky with a magnitude of $\sim$12 km s$^{-1}$, and that the outflowing material is mainly located on the thin edges of a cone with an opening angle of 30$^{\circ}$. The P06 authors did not detect rotation in the outflow of HH30, and set an upper limit of 1 km s$^{-1}$ at 200 au from the jet axis.

\smallskip

In this paper, we report the first Atacama Large Millimeter/Submillimeter Array (ALMA) band 6 CO and continuum observations at $\sim$0.25'' angular resolution (or $\sim$35 au) of the HH30 system, aimed at constraining the outflow features. We detail our observations and data reduction in Sect.~\ref{s:obs} and develop our analysis of the continuum and of the $^{13}$CO and $^{12}$CO emission lines in Sect.~\ref{s:result}. Section~\ref{s:analysis} presents our detailed analysis of the $^{12}$CO emission. The Sect.~\ref{s:discu} discusses the origin of the outflow of HH30 and give constraints on the central binary system. We summarize our conclusions in Sect.~\ref{s:concl}.

\section{Observations and data reduction}
\label{s:obs}

The characteristics of our lines and continuum observations are detailed below. The resulting beam sizes and sensitivities are summarized in Table~\ref{t:obs}. The $^{12}$CO($J$=2$\rightarrow$1),  $^{13}$CO($J$=2$\rightarrow$1), and C$^{18}$O($J$=2 $\rightarrow$1) emission lines plus the continuum emissions at 1.28 mm and 1.38 mm of HH30 were observed using the Band 6 of ALMA (211-275GHz) at the phase centre $\alpha$(J2000) = 04:31:37 and $\delta$(J2000) =18:12:24. The data were taken using the cycle 2 semi-extended configuration of ALMA with baselines ranging from 13 m to 1570 m. Two tracks were performed on  July 19, 2015 and one on July 21, 2015. Each track lasted $\sim$40 minutes. The Band 6 data contained three spectral windows of 117.2 MHz bandwidth each in 960 channels that were tuned at 219.563 GHz, 220.379 GHz, and 230.546 GHz to simultaneously cover the C$^{18}$O($J$=2 $\rightarrow$1),  $^{13}$CO($J$=2$\rightarrow$1), and  $^{12}$CO($J$=2 $\rightarrow$1), respectively. Two additional spectral windows of 2\,GHz bandwidth in 128 channels centred at 217.044 GHz and 234.010 GHz were dedicated to the detection of the continuum emission from the HH30 dust disk. 

The data were reduced using the common astronomy software application \citep[hereafter CASA; see][]{mcmullin06}. We performed an initial correction for rapid atmospheric variations at each antenna using water vapor radiometer data and corrected for the time and frequency dependence of the system temperatures. One of the tracks from July 19, 2015 (uid\_\_\_A002\_Xa5df2c\_X9030) was put aside due to irregular phase drifts over short (minutes) timescales. Bandpass and flux calibrations were performed on the quasar \object{J0423-0120}. Quasar \object{J0510+1800} was used in both remaining tracks to calibrate the time variation of the complex gains. Based on the dispersion between the fluxes derived for the phase calibrator in each observing session, we estimate the absolute flux calibration to be accurate within $\sim$10\%. Owing to residual inconsistency in the phase calibration between the two tracks, we used the continuum spectral windows to derive accurate phase centre and re-project the data cubes before merging. Imaging was carried out using the cleaning method HOGBOM of the GILDAS\footnote{See the following web page for details: https://www.iram.fr/IRAMFR/GILDAS/.} package. With a ROBUST weighting and a parameter of 0.56\footnote{Therefore giving slightly more weight to the longest baselines than with natural weighting.} the synthesized beam has in average a size of 0.26$^{\prime\prime}$ $\times$0.19$^{\prime\prime}$ at a position angle (PA) of $\simeq$29$^\circ$ (see Table~\ref{t:obs} for details on each spectral window). We merged the two continuum emission spectral windows at 217.044 GHz and 234.010 GHz. Assuming that the continuum spectrum is accurately described by $S(\nu) \propto \nu^{2.2}$ in the disk, as derived by P06, the merged continuum emission results in a typical 225.964 GHz (i.e. 1.33 mm) continuum emission. The resulting continuum emission (see Section~\ref{ss:cont} and Fig.~\ref{f:cont}) has a root mean square (rms) noise level of 21.74 $\mu$Jy/beam (see Table~\ref{t:obs}). We used the 234.010 GHz continuum emission to subtract from the $^{12}$CO(2-1) emission and the 217.044 GHz continuum emission to subtract from the $^{13}$CO(2-1) emission. The channel spacing of 122 kHz in the molecular line spectral windows resulted in a native velocity resolution of 0.16 km~s$^{-1}$ and 0.17 km~s$^{-1}$ in $^{12}$CO(2-1) and $^{13}$CO(2-1), respectively, which we degraded to 0.3 km~s$^{-1}$ to detect weak emission. The $^{12}$CO(2-1) and $^{13}$CO(2-1) emission lines display a rms noise level of  1.9~mJy/beam and 2.15~mJy/beam per channel, respectively.


\section{Results}
\label{s:result}

In this paper, we focus on the $^{12}$CO data that primarily trace the molecular outflow/cavity. To constrain more accurately the  source position, V$_{\rm lsr}$, and disk inclination, we also report the main observational results for the 1.3~mm continuum and $^{13}$CO emissions. However, a detailed analysis of the disk structure is beyond the scope of this paper and will be conducted in a forthcoming publication.

\begin{table*}
\caption{Observational set-up of the ALMA-cycle 2 data set}
\label{t:obs}
\begin{center}
\addtolength{\tabcolsep}{+0pt}
\begin{tabular}{lccccc}
\hline
\hline                                                                                                                                                                                    \\ 
Spectral window         & $^{12}$CO(2--1)                & $^{13}$CO(2--1)                & \multicolumn{3}{c}{Continuum}                                                                     \\
                        &                                &                                &     1.28 mm                     & 1.38 mm                     & 1.33 mm$^a$                   \\        
\hline                                                                                                                                                                                        
Frequency               & 230.538 GHz                    & 220.398 GHz                    & 234.006 GHz$^b$                 & 217.040 GHz$^b$             & 225.204 GHz$^b$               \\ 
Bandwidth               & 117 MHz                        & 117 MHz                        & 2 GHz                           & 2 GHz                       & 4 GHz                         \\
Native channel width    & 122 kHz                        & 122 kHz                        & 15.6 MHz                        & 15.6 MHz                    & -                             \\
Primary beam            & 25.3\arcsec                    & 26.4\arcsec                    & 24.9\arcsec                     & 24.9\arcsec                 & -                             \\
Synthesized beam        & $0\farcs24\times0\farcs19$     & $0\farcs25\times0\farcs20$     & $0\farcs25\times0\farcs18$      & $0\farcs28\times0\farcs19$  & $0\farcs26\times0\farcs18$    \\
Beam PA     & 26.8$^{\circ}$                 & 31.2$^{\circ}$                 & 27.2$^{\circ}$                  & 31.2$^{\circ}$              & 29.7$^{\circ}$                \\ 
\hline
Rms noise level$^c$     & 2.0 mJy/beam                   & 2.3 mJy/beam                   & 30.96 $\mu$Jy/beam              & 29.15 $\mu$Jy/beam          & 21.74 $\mu$Jy/beam            \\
Flux$^d$                & 15.4 Jy.km s$^{-1}$            & 0.88 Jy.km s$^{-1}$            & 24.53 mJy                       & 20.85 mJy                   & 22.30 mJy                      \\
\hline
\end{tabular}
\end{center}
($^a$): The continuum at 1.33~mm results from the merging of the 1.28~mm and 1.38~mm continuum bands.\\
($^b$): The mean frequency was calculated assuming a S($\nu$)\,$\propto \nu^{-2.2}$ emission spectra accurately describes the interstellar medium spectral energy distribution slope in frequency range considered. \\
($^c$): The noise level of the CO transition lines are given per channel, and the channels have a width of 0.3 km s$^{-1}$. \\
($^d$): The continuum and $^{12}$CO(2--1) fluxes are integrated within the area given by the 5$\sigma$ level. The $^{13}$CO(2--1) flux is integrated within the area given by the 3$\sigma$ level.
\end{table*}

\subsection{Continuum emission}
\label{ss:cont}

\begin{figure*}[h!]
\centerline{
\includegraphics[trim = 0cm 15cm 0cm 0cm ,width=1.1\textwidth]{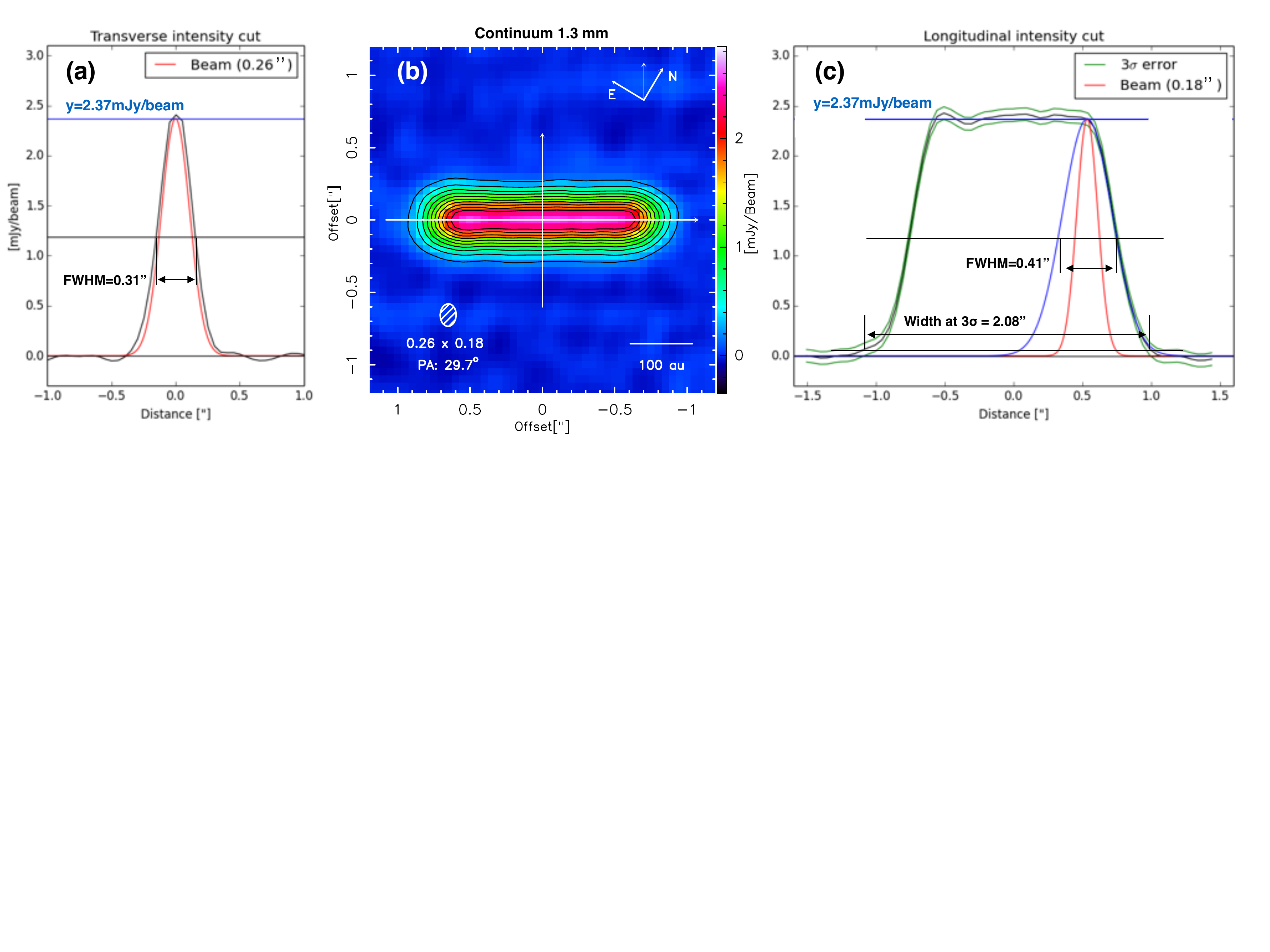}}
\caption{HH30 1.33 mm continuum emission. \textbf{Left:} Intensity cut across the minor axis of the disk (black curve) compared to the beam (red curve). \textbf{Middle:} Continuum emission map at 1.33 mm rotated by 31$^\circ$ and centred at $\alpha$(J2000) = 04:31:37.5 and $\delta$(J2000) =18:12:23.8. The contours start at 10$\sigma$ with 10$\sigma$ steps. The 1$\sigma$ noise level is of 21.74 $\mu$Jy/beam. The white arrow localizes the intensity cut along the disk shown on the right panel. The orientation is shown in the top right corner. \textbf{Right:} Intensity cut along the major axis of the disk. The blue Gaussian adjusts the decrease in flux beyond r=75 au, which is broader than the beam (red curve). The green curves represent the $\pm$3$\sigma$  variation of the longitudinal intensity profile.}
\label{f:cont}
\end{figure*}

Figure~\ref{f:cont}b shows the 1.33 mm continuum emission of HH30. With a signal-to-noise ratio of $\sim$120 the continuum emission in the disk of HH30 is clearly detected. The disk axis has a PA of $31.2^\circ \pm0.1$. The orientation is consistent with the dark lane seen in the HST images (disk axis at PA 32.2$^\circ \pm$ 1.0, \citealt{burrows96}) and with the previous millimetre study of P06 and \cite{guilloteau08}. It also remarkably agrees with the jet axis PA close to the star of 31.3$^{\circ}$ \citep{burrows96}.

Figures~\ref{f:cont}a\&c show the intensity cuts across and along the disk, respectively. The intensity cut across the disk has a homogeneous full width at half maximum (FWHM) of $\sim$0.31$^{\prime\prime}$ at all radii, close to the theoretical beam size in that direction, 0.26$^{\prime\prime}$. The deconvolved vertical size, calculated as $\sqrt{0.31''^2-0.26''^2}$, gives an upper limit for the disk vertical FWHM of $\sim$0.17$^{\prime\prime}$=24~au. Along the disk plane, the flux of the longitudinal cut varies by less than 5\%. Considering a possible flux variation of $\pm 3\sigma$ (green curves on Fig.~\ref{f:cont}c), where $\sigma$ is the rms of the continuum map, the plateau is compatible with a constant flux of $\sim$2.4 mJy/beam. This contradicts the previous study of \cite{guilloteau08}, which found the disk of HH30 to be truncated at an inner radius of 0.26$^{\prime\prime}$ (or 37 au) from their $\sim$0.44$^{\prime\prime}$ PdBI observations. 
The longitudinal intensity cut shows a sharp decay at the edges of the disk, resolved by our data set, and betraying a rapid fall-off in brightness for radii $\ge$~0.55$^{\prime\prime}$=75~au.
We estimate the diameter of the disk to be 2.08$^{\prime\prime}$, calculated as its full width above the 3$\sigma$ level. With the adopted distance of the source of 140 pc, this corresponds to a physical radius of 145~au, which is slightly larger than the radius of 130~au estimated by \citet{guilloteau08}. The disk appears much smaller in 1.3~mm continuum emission than in scattered light in optical images, where it extends out to a radius of 250 au \citep{burrows96}.

 Assuming that the continuum emission comes from a thin layer of dust in the equatorial plane of the disk, the aspect ratio derived from the longitudinal and transverse cuts provides a lower limit to the disk inclination of $i > 84.8^{\circ}$. This value is compatible with the disk inclinations of 82.5-84$^{\circ}$ inferred from the brightness asymmetry in the optical lobes \citep{burrows96,cotera01, wood02} and with the disk inclination of 81$^{\circ}$ $\pm$ 3$^{\circ}$ derived by P06 from fitting the $^{13}$CO and continuum emissions observed at the PdBI.

 The resulting flux density of 22.30$\pm$0.05 mJy at 1.33 mm agrees with previous measurements made at Owens Valley Radio Telescope \citep{stapelfeldt99} and PdBI (P06). We do not attempt to derive a mass estimate from this flux since the assumption of optically thin emission is likely not valid in HH30 owing to its very close to edge-on geometry.

\subsection{$^{13}$CO emission \& V$_{\rm lsr}$}
\label{ss:13co}

\begin{figure*}[h!]
\centerline{
\includegraphics[width=1\textwidth]{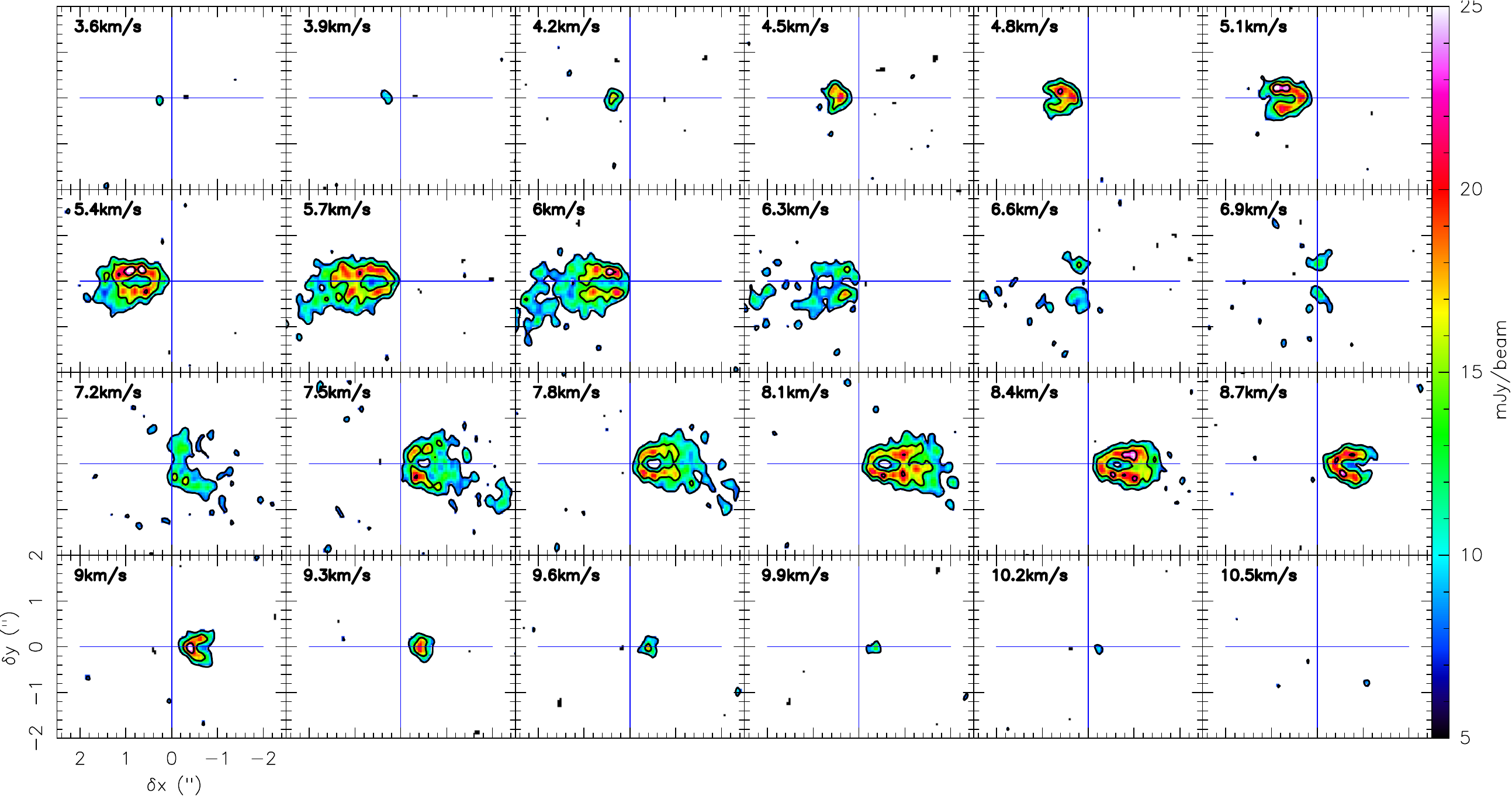}}
\caption{Channel maps of the $^{13}$CO(2-1) emission line of HH30. The contours start at 3$\sigma$ with 3$\sigma$ steps with $\sigma=2.30$ mJy beam$^{-1}$ (or 1.25 K). The channel velocity is indicated at the top in km~s$^{-1}$. The blue cross locates the central position derived from the continuum.}
\label{f:chan13}
\end{figure*}

Figure~\ref{f:chan13} shows the $^{13}$CO(2-1) channel maps. The $^{13}$CO emission is centred on the continuum emission. The disk is detected from $\sim$3.2 km s$^{-1}$ to 10.1 km s$^{-1}$. The two faces of the disk are visible at intermediate velocities from 5 km s$^{-1}$ to 6.2 km s$^{-1}$ and from 7.7 km s$^{-1}$ to 8.9 km s$^{-1}$ on both sides of the mid-plane.

\begin{figure*}
\centerline{
\subfloat{\includegraphics[width=0.5\textwidth]{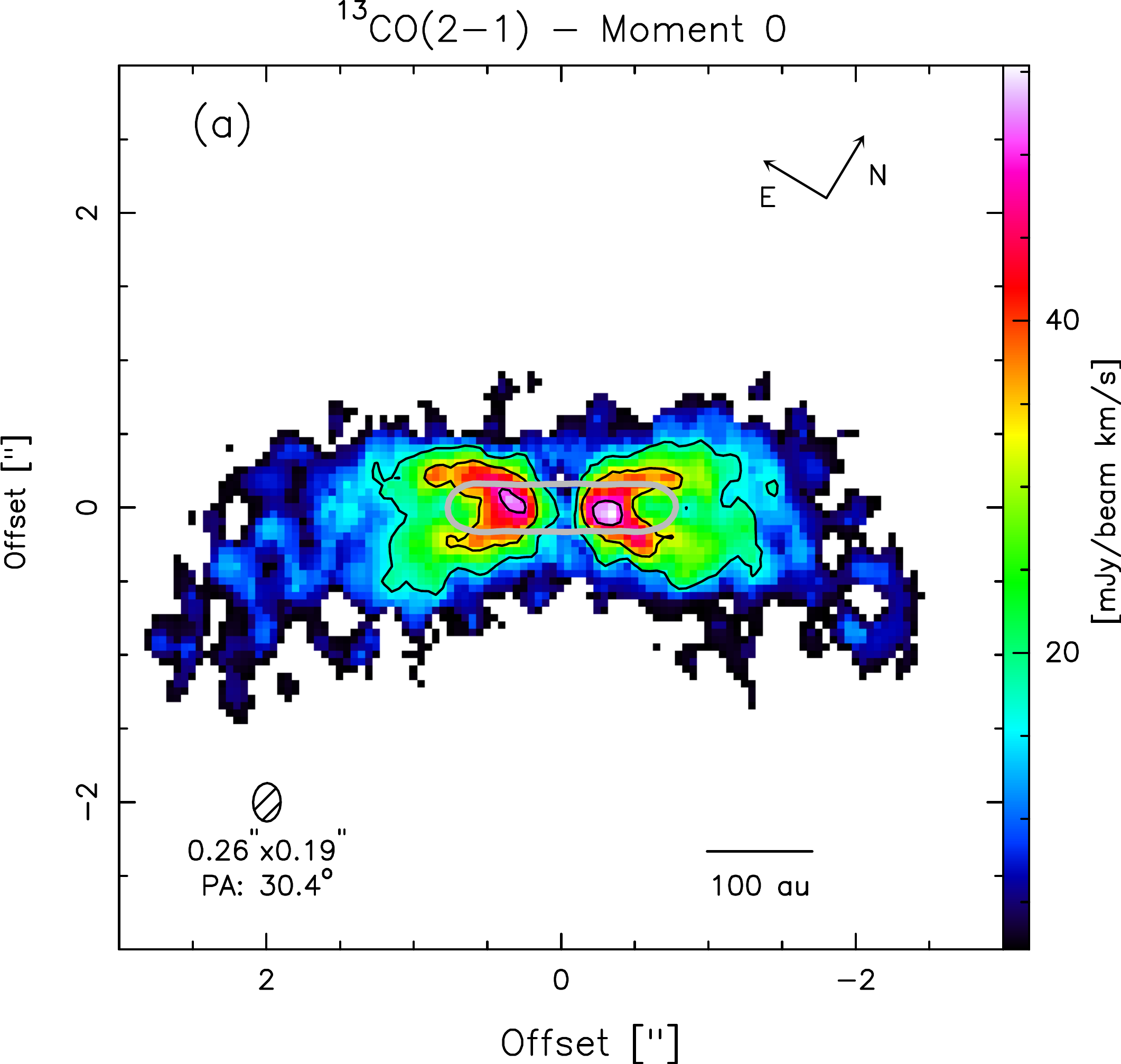}}
\subfloat{\includegraphics[width=0.5\textwidth]{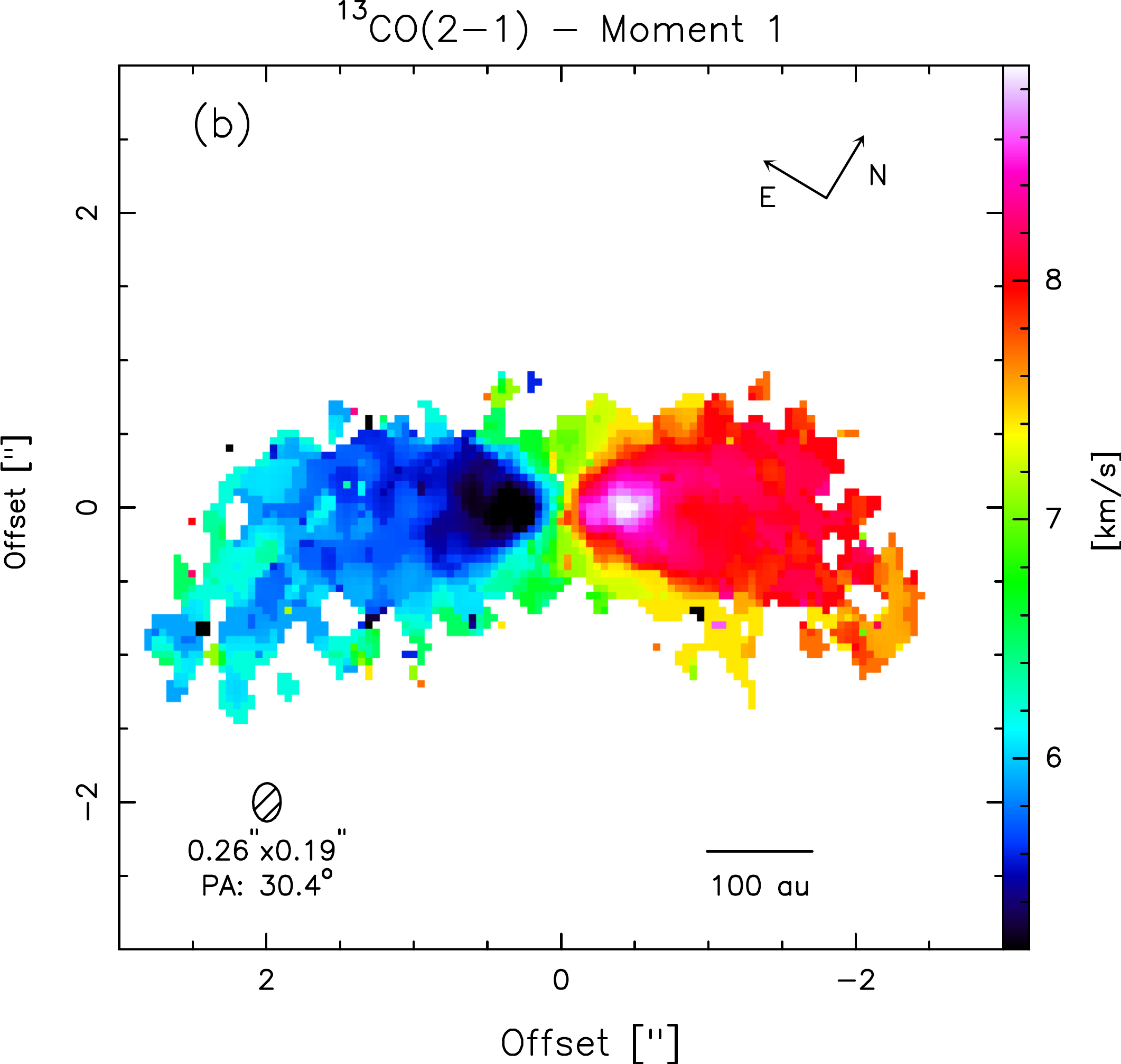}}}
\caption{\textbf{Left:} Moment zero of the $^{13}$CO(2-1) emission line integrated from 3.4 km s$^{-1}$ to 11.1 km s$^{-1}$. The contours start at 3$\sigma$ with 3$\sigma$ steps with $\sigma=$ 5.55 mJy/beam km s$^{-1}$. The contour in grey represents the level at 50$\sigma$ of the continuum emission (see Fig.~\ref{f:cont}). \textbf{Right:} First moment map of the $^{13}$CO(2-1) emission line. The beam is shown in the bottom left corner and N-E orientation as indicated in the top right corner.}
\label{f:13co}
\end{figure*}

Figure~\ref{f:13co}a shows the moment 0 of the $^{13}$CO(2--1) emission line integrated over the velocity range [3.4 km s$^{-1}$, 11.1 km s$^{-1}$]. It has an integrated flux above the 3$\sigma$ level of 0.88 Jy km s$^{-1}$. The $^{13}$CO emission extends further out than the continuum emission; there is a detection above 3$\sigma$ up to a radius of $\sim$1.3$^{\prime\prime}$ (or 182~au) that is comparable to the radial extension of the optical nebulosity. We note an apparent lack of $^{13}$CO emission in the central part of the disk. We suggest that this is due to an obscuration effect of the inner $^{13}$CO disk emission by the outer continuum emission resulting from the very close to edge-on configuration. 
A radiative transfer disk model including both gas and dust would be necessary to confirm this interpretation but is outside the scope of the present paper. The clear detection of both faces of the disk with a characteristic butterfly pattern confirms that the disk is seen nearly edge-on (see Sect~\ref{ss:cont}). The channel maps show a fairly symmetric emission in $^{13}$CO with respect to the plane of the disk at all velocities. The flux ratio between the emission from the two disk faces is compatible with $\sim$1 in all channel maps with a standard deviation of 15\%. Figure~\ref{f:13co-res} further shows the $^{13}$CO residual channel maps, after subtracting the symmetric bottom hemisphere disk emission from the top hemisphere emission. It shows that very few positive (or negative) emission remains after subtraction, which confirms that the emission in $^{13}$CO(2-1) is very symmetrical with respect to the mid-plane.

\begin{figure}
\centerline{
\includegraphics[width=0.5\textwidth]{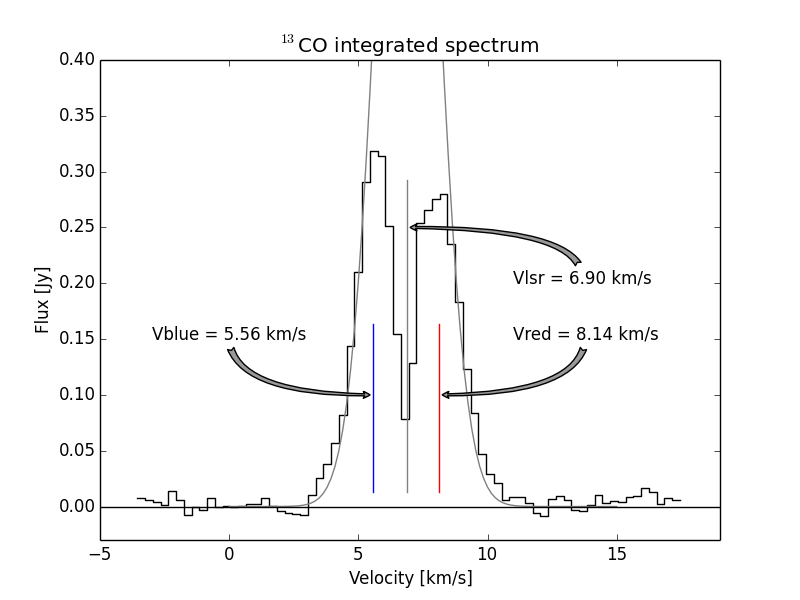}}
\caption{Integrated spectrum of the $^{13}$CO(2-1) emission line over the area above the 3$\sigma$ level shown in left panel of Fig.~ \ref{f:13co}.}
\label{f:13cospec}
\end{figure}

Figure~\ref{f:13cospec} shows the integrated spectrum of the $^{13}$CO(2--1) emission line summed over the area defined by the 3$\sigma$ contour on the moment 0 map (see Fig.~\ref{f:13co}a). The integrated line profile is double peaked with symmetric wings, exhibiting the classical profile of rotating Keplerian disks in cTTS \citep{duvert00,guilloteau94,louvet16b}. The right panel of Fig.~\ref{f:13co} shows the moment 1 map of the $^{13}$CO(2-1) emission. The emission arising from the south-eastern part of the disk is blue-shifted, while the north-western part of the disk is red-shifted confirming its rotation.  

 The $^{13}$CO emission line profile is best fitted with two 1D Gaussian with the same peak flux of about 0.30 Jy and the same FWHM of 1.8~$\pm$~0.3 \kms. The two Gaussian are separated by 2.60 $\pm$ 0.05 \kms. To determine the $V_{\rm lsr}$ of \object{HH30}, we fit a single Gaussian component to the high-velocity wings of the $^{13}$ CO profile (see Fig.~\ref{f:13cospec}). We derive a central velocity of $V_{\rm lsr} = 6.9$ $\pm$ 0.1 \kms~ in $^{13}$CO, which is slightly different from the value previously derived by P06 in $^{13}$CO (7.25 $\pm$ 0.04 \kms).
 
 Our ALMA observations therefore unambiguously confirm that the $^{13}$CO emission is arising from the HH30 rotating disk, in accordance with the previous results of P06. No contribution from the outflow is detected in $^{13}$CO. A detailed analysis of the $^{13}$CO emission from the disk goes beyond the scope of this paper and will be conducted in a forthcoming publication. 
   
\subsection{$^{12}$CO emission}
\label{ss:12co}

\begin{figure*}
\subfloat{\includegraphics[width=1\textwidth]{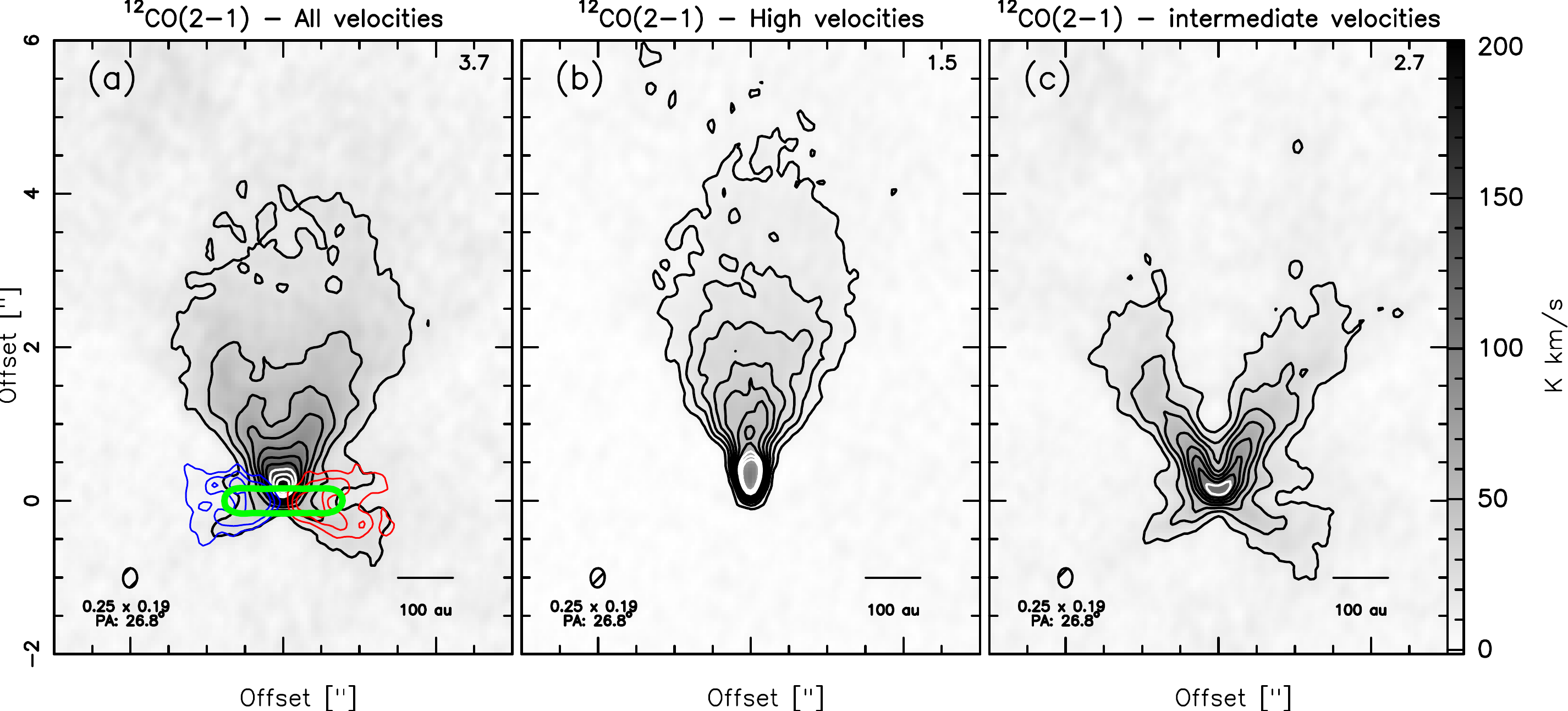}}\\
\subfloat{\includegraphics[width=1\textwidth]{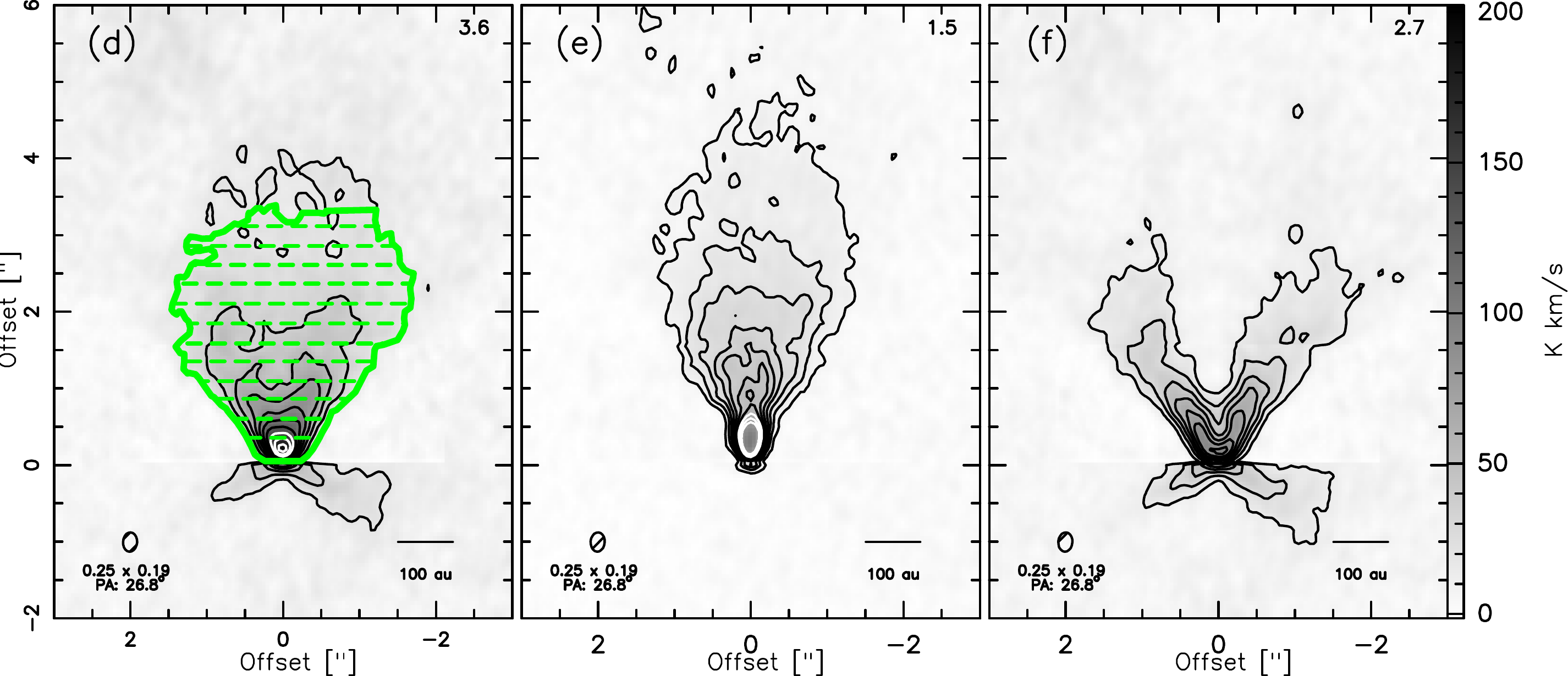}}
\caption{ \textbf{(a):} Moment 0 of the $^{12}$CO(2-1) emission, integrated over the full range of emission from -0.1 km s$^{-1}$ to 14 km s$^{-1}$. The blue and red contours highlight the blue-shifted and red-shifted $^{13}$CO(2-1) emission line arising from the disk.\textbf{(b):} $^{12}$CO emission integrated at high velocities from -0.1 km~s$^{-1}$ to 2.3 km~s$^{-1}$ and from 11.3 km~s$^{-1}$ to 14 km~s$^{-1}$. At these velocities, only the outflow contributes to the $^{12}$CO emission. \textbf{(c)} $^{12}$CO emission integrated over intermediate velocities from 2.6 km~s$^{-1}$ to 11 km~s$^{-1}$. At those velocities, the $^{12}$CO emission is a mixture of emissions arising from the outflow and from the disk. \textbf{Bottom panels:} Same as the top panels, after subtraction of the emission arising from the disk in the northern hemisphere (see Sect.~\ref{ss:sous}). Panel \textbf{(d):} The thick green contour defines the area used to derive the mass of the outflow (see Sect.~\ref{ss:massf}). Together with the green horizontal lines, it defines the 13 regions used to construct the temperature brightness profile along the flow shown in Fig.~\ref{f:massf}. In all panels the contours start at 5$\sigma$ with 5$\sigma$ steps. The 1$\sigma$ value is indicated on the top right corner of each panel in the unit K km s$^{-1}$.}
\label{f:12co}
\end{figure*}

\begin{figure*}[h!]
\centerline{
\includegraphics[width=1\textwidth]{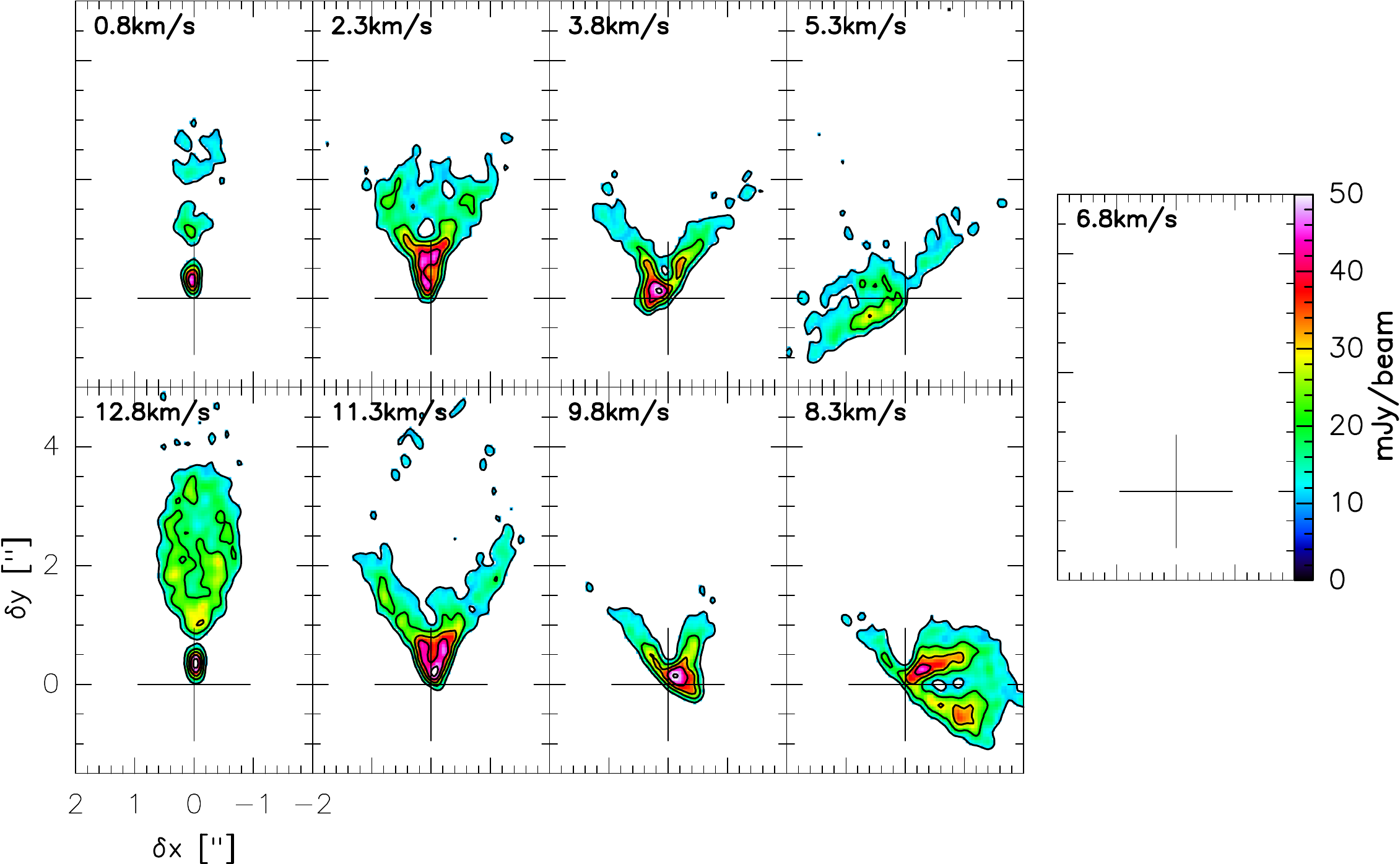}}
\caption{Selected representative channel maps of the $^{12}$CO(2-1) emission line in HH30. The contours start at 5$\sigma$ with 5$\sigma$ steps with $\sigma=2.0$ mJy beam$^{-1}$ (or 0.8 K). The channel velocity is indicated in the top left corner in km~s$^{-1}$. The black cross locates the centre position derived from the continuum. Source V$_{lsr}$=6.9$\pm$0.1\kms.}
\label{f:chan12-sous}
\end{figure*}

Figure~\ref{f:12co} shows the $^{12}$CO(2-1) total intensity map (top left panel), the intensity map integrated over high velocities (top middle panel), and the intensity map integrated over intermediate velocities (top right panel). It clearly illustrates that the $^{12}$CO emission line originates from two components: from the circumstellar disk, on the one hand, and from the outflow, on the other hand. The $^{12}$CO emission arising from the disk extends up to $r=1.1^{\prime\prime}$, as the $^{13}$CO emission. Low-level emission from the disk surface seems more extended towards the south-west in $^{12}$CO. A similar effect, but in the opposite sense, is present in scattered light and has an enhanced brightness to the south-east in optical images of  \cite{burrows96}. Such effects have been attributed to non-axisymmetric illumination of the outer disk. Our $^{12}$CO data show no difference in kinematics between the south-east and the south-west emissions; this is consistent with this interpretation.
Figure~\ref{f:chan12-sous} shows selected channel maps of the $^{12}$CO(2-1) emission while all channel maps are shown in Fig.~\ref{f:chan12}. 
The $^{12}$CO(2-1) emission in HH30 spreads from -0.1 km~s$^{-1}$ to 14 km~ s$^{-1}$, i.e. in a much wider interval than the $^{13}$CO(2-1) emission line. Both the channel maps and integrated intensity maps clearly reveal the monopolar CO outflow towards the north, perpendicular to the disk plane. No signature of CO outflow is detected towards the south down to our sensitivity level.

At high velocities $v\geq 11.3$ km~s$^{-1}$ and $v\leq 2.3$ km~s$^{-1}$ the $^{12}$CO emission can be attributed to the outflow without ambiguities. The north-eastern cavity/outflow is detected up to $\sim$4$^{\prime\prime}$ away from the central protostar in the high-velocity red-shifted channels. Considering the distance of HH30 (140 pc) and the inclination of the disk ($>84.8^\circ$, see Sect.~\ref{ss:cont}), this corresponds to a physical scale of $\sim$560 au. At intermediate velocities 7.4 km~s$^{-1}$ $\leq v \leq$ 11 km~s$^{-1}$ and 2.6 km~s$^{-1}$ $\leq v \leq$ 5.3 km~s$^{-1}$ the $^{12}$CO emission is a mixture of emissions arising from both the disk and outflow. In a similar way as the $^{13}$CO emission (see Sect.~\ref{ss:13co}), both faces of the disk can be distinguished at intermediate velocities from 3.5 km~s$^{-1}$ to 5 km~s$^{-1}$ and from 7.7 km~s$^{-1}$ to 10.4 km~s$^{-1}$. From 5.6 km~s$^{-1}$ to 7.4 km~s$^{-1}$, i.e. around the systemic velocity of HH30 (v$_{\rm lsr}\sim$ 6.9 km~s$^{-1}$; see Sect.~\ref{ss:13co}) the confusion with the molecular cloud is so important that the disk remains undetected. 
 
The channel maps are almost exactly symmetric with respect to the systemic velocity of HH30, which supports the outflow to be seen nearly in the plane of the sky. A brightness asymmetry between front and back side of the cavity is clearly apparent: high-velocity red-shifted emission is detected out to 4$^{\prime\prime}$, which is significantly farther than high-velocity blue-shifted emission only detected out to 2$^{\prime\prime}$.  
 
The V-shape morphology of the $^{12}$CO outflow emission at intermediate velocities and a more collimated morphology at high velocities is globally consistent with the previous observations of P06 that were interpreted as signatures of a conical outflow. Indeed, if the gas flows at a constant velocity along the surface of a cone observed close to edge-on, the high-velocity emission traces matter located at the front/back sides of the cone, while the low-velocity emission traces the edges of the cone. Thus high-velocity channel maps appear more collimated while low-velocity channel maps more closely outline the opening angle of the cone \citep[see Fig.~11 in][]{pety06}. This general trend is followed by our ALMA observations. The interpretation of a conical outflow is also supported by the hollow ellipse morphology of transverse position-velocity (pv) diagrams, which are discussed below.

At intermediate to high velocities, especially in the red-shifted part of the flow (see e.g. channel maps at 11.6 \kms to 12.5 km~s$^{-1}$ in Fig~A.1), the emission in channel maps form closed elliptical structures that might suggest an expanding bubble geometry. Similar behaviour has been observed for example at the base of the DG~Tau micro-jet in optical emission lines with HST \citep{bacciotti00}. However, such elliptical channel maps can also be produced in a cone when the sideways expansion velocity (Vr) decreases beyond some distance to the source. We argue below from the analysis of the longitudinal pv diagrams that this is what occurs in HH30.

The channel maps at high velocity seem to indicate an apparent change of morphology close to the source at y $\leq$ 0.5$^{\prime\prime}$ where the emission appears more cylindrical. This effect results from the combination of projection and beam convolution effects because the base of the cone is not fully resolved. We come back to this issue in section~\ref{s:analysis}.

\begin{figure*}
\centerline{
\subfloat{\includegraphics[width=1\textwidth]{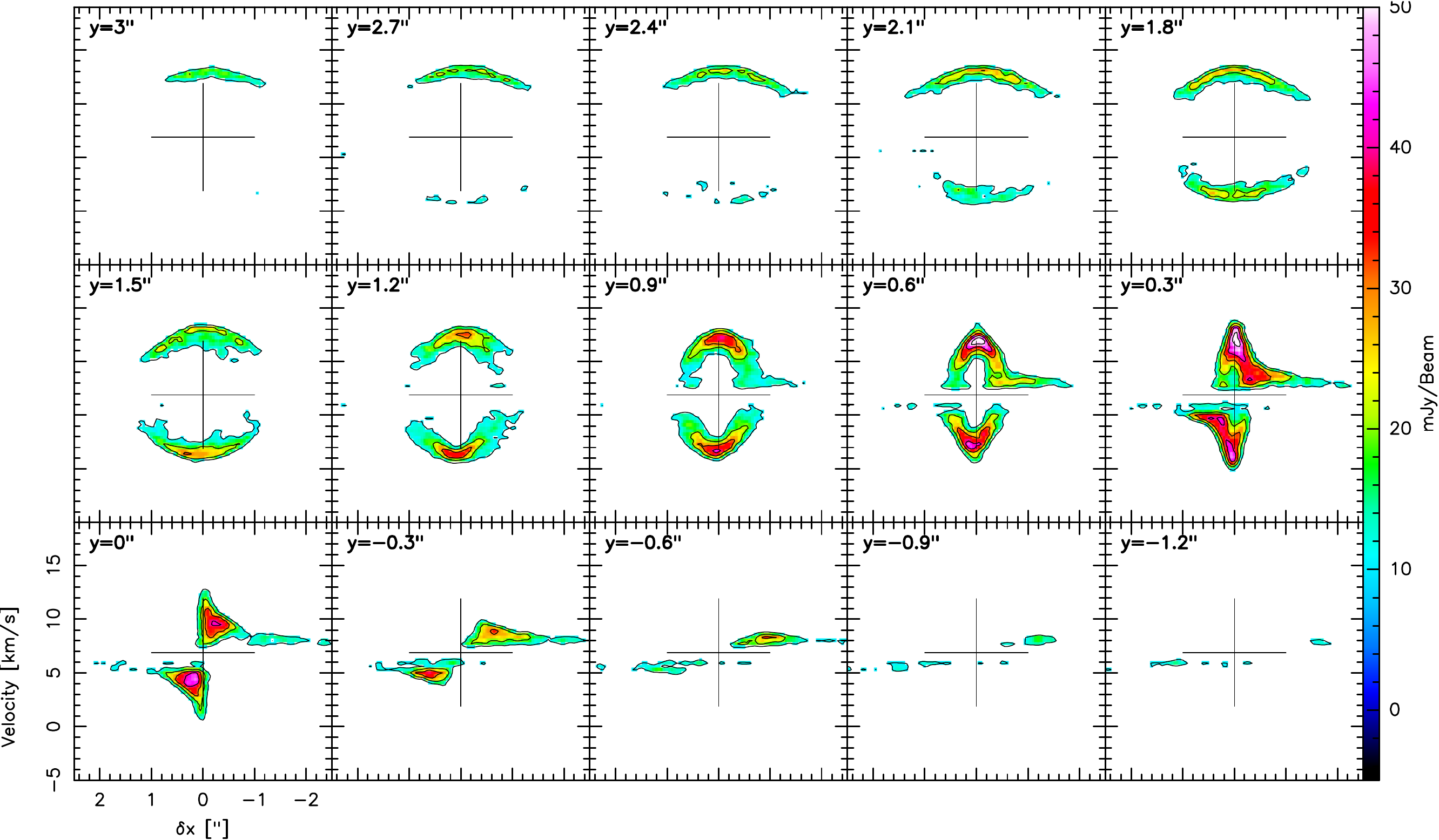}}}
\caption{Transverse pv diagrams of the $^{12}$CO(2-1) emission, with pseudo-slit parallel to the disk main axis, from y=+3$^{''}$ on the top left to y=-1.2$^{''}$ on the bottom right. Signatures of the Keplerian disk dominate for $y<0.3^{\prime\prime}$ while elliptical signatures of the CO cavity/outflow appear at $y > 0.3^{\prime\prime}$. The black cross shows the central $x=0^{\prime\prime}$ position and source $v_{\rm lsr}$ at 6.9 km s$^{-1}$. The contour levels start at 5$\sigma$ with 5$\sigma$ steps with $\sigma=$2.0 mJy/beam.}
\label{f:12co-pvxy}
\end{figure*}

Figure~\ref{f:12co-pvxy} shows the $^{12}$CO pv diagrams transverse to the flow axis (i.e. slices parallel to the disk equatorial plane) from $y=-1.2''$ (i.e. southern face of the disk) and up to $y=3''$ in the outflow. The Figure~\ref{fa:12co-pvxy} further shows the pv diagrams up to $5''$ For negative values of $y$, we clearly see the disk rotation signature with red-shifted emission on the north-western side and blue-shifted emission on the south-eastern side, which agrees with the $^{13}$CO data (see Sect.~ \ref{ss:13co}). The plot at $y=0$ (i.e. in the plane of the disk) show pv diagrams with a quasi perfect point symmetry, as expected from a disk in Keplerian rotation. The contribution of the disk is visible up to $y=0.9''$ in the red-shifted channels at intermediate velocities while it stops contributing at $y \ge 0.72''$ in the blue-shifted channels. From $y=0.36^{\prime\prime}$ and up to $y\sim$2$^{\prime\prime}$, the pv diagrams show an additional elliptical component. Such an elliptical pv diagram is expected if emission arises in a hollow conical shell: the two peaks at $V-V_{\rm lsr} \simeq \pm 5$ km~s$^{-1}$ close to the flow axis trace projected emission from the back and front sides of the cone while faint emission close to $V_{\rm lsr}$ detected at larger transverse distances from the flow axis trace the sides of the cone. Both faces of the cone are clearly detected out to y=2$^{\prime\prime}$, where emission from the blue-shifted (front) side  vanishes into the noise level while red-shifted emission is detected farther out in agreement with the channel maps. 

\begin{figure}
\centerline{
\includegraphics[trim = 0cm 0cm 0cm 0cm, width=0.5\textwidth]{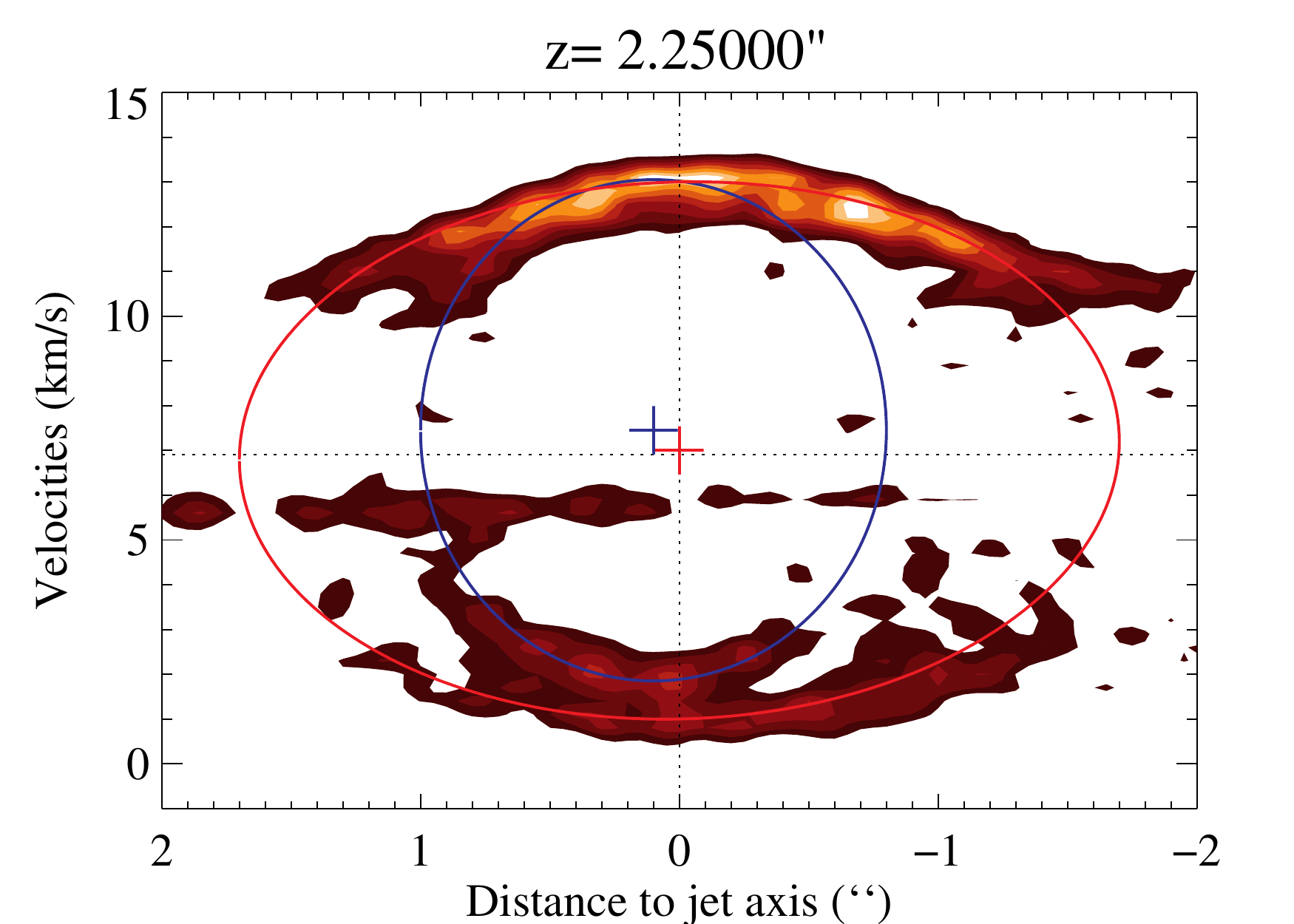}}
\caption{High contrast pv diagram of the slice at $y=2.25^{\prime\prime}$ presented in Fig.\ref{f:12co-pvxy}. 
The ellipses in blue and red highlight the two possible shells visible at this position. The two crosses show their respective centre.}
\label{f:pvxy22}
\end{figure}

From $y\gtrsim2^{\prime\prime}$, a second ellipse is partially visible in the transverse pv diagrams out to $y \simeq 5^{\prime\prime}$ (see Fig.~\ref{fa:12co-pvxy}). It is detected mostly at blue-shifted velocities for distances $y\leq3.5^{\prime\prime}$ and at red-shifted velocities for distances beyond. The Figure~\ref{f:pvxy22} highlights the two components at the altitude $y=2.25''$ above the disk plane. The radius of the second component is smaller than that of the main outer conical cavity. Therefore, this inner elliptical emission betrays an inner-shell of material inside the outer conical cavity. Its maximum projected line-of-sight velocity is always smaller than that of the outer cavity, and its centroid velocity appears  red-shifted. We come back to this component and its relation with the outer cavity in the following section.

The longitudinal pv diagrams, which have slices perpendicular to the disk equatorial plane from $x=1.5^{\prime\prime}$ to $x=-1.5^{\prime\prime}$, are given in Appendix~\ref{fa:12co-pvyx}. All pv diagrams show a lack of emission near the systemic velocity at $v=6.9$~km~s$^{-1}$ because of the filtering of large-scale CO emission by the ALMA interferometer. We clearly see the two faces of the disk from $x=-1^{\prime\prime}$ to $x=+1^{\prime\prime}$ in agreement with the channel maps, and the disk radial velocity decreases with increasing radius as expected from a disk in Keplerian rotation. The outflow contributes to the $^{12}$CO emission from $x=+1.2^{\prime\prime}$ to $x=-1.2^{\prime\prime}$. It is composed of two dominating velocities at $V-V_{\rm lsr} \simeq \pm 5$ km~s$^{-1}$ betraying a nearly constant radial velocity with altitude for the front and back sides of the cavity. On the contrary, for a spherical, isotropically expanding bubble of radius R$_0$ and expansion speed V$_0$, we would expect the maximum line-of-sight velocity at each height z to vary linearly with the projected radius of the bubble; the latitude $\theta$ from the bubble equator is defined as sin($\theta$) = (z-z$_0$)/R$_0$, where z$_0$ is the centre of the bubble,  V$_{\rm max}$ = V$_0$ cos($\theta$), and Rmax = R$_0$ cos($\theta$), hence V$_{\rm max}$(z) = (V$_0$/R$_0$) x R$_{\rm max}$(z). This behaviour is clearly not observed in this case. Between y=0.6$^{\prime\prime}$ and y=1.5$^{\prime\prime}$, the radius of the cavity increases by a factor $\simeq$ 2 while the maximum line-of-sight velocity stays roughly constant (see Fig.~\ref{f:12co-pvxy} ). A slight decrease in line-of-sight velocity is observed beyond y=3$^{\prime\prime}$ for the red-shifted side (see the panels at x=$\pm$~0.3$^{\prime\prime}$ in Fig~A.3). This slight decrease is responsible for the closing back of the contours towards the axis, creating the elliptical shapes seen in the channel maps at high red-shifted velocities.

For x$\in[-0.2'';0.2'']$ and at an altitude of y$\sim$0.25$''$ above the disk mid-plane there is a high-velocity ($V-V_{\rm lsr} \simeq \pm 7$ km~s$^{-1}$) component that seems disconnected from the disk and outflow components (see orange ellipses on Fig.~\ref{fa:12co-pvyx}). This component is only detected on axis. Higher angular resolution observations is required to determine if this component is related to the base of the conical cavity or traces a high-velocity emission knot related to the inner jet variability. We come back to this issue in Sect~\ref{ss:entrenement}.

In summary, the morphology and kinematics of the HH30 $^{12}$CO emission retrieved by our ALMA data  agree with the conical outflow interpretation previously derived by P06 out to y=2$^{\prime\prime}$= 280~au where both faces of the cone are clearly detected. Beyond these distances we detect signatures of at least one additional inner cavity/shell. We also detect a possible on axis high-velocity knot at distances y$\sim+0.25''$ above the disk mid-plane. Additionally, both the channel maps and the kinematic behaviour of the southern $^{12}$CO emission appear dominated by the disk atmosphere.

\section{Analysis}
\label{s:analysis}

We develop in this section a simple geometrical model to derive the kinematics and morphology of the CO cavity from the transverse pv diagrams. We first accurately isolate the CO outflow emission from the disk contribution and estimate its mass and brightness temperature distribution. We detail our geometrical modelling and fitting procedure. We then discuss the results, in particular regarding rotation and wiggling signatures.

\subsection{Subtraction of the $^{12}$CO emission arising from the disk}
\label{ss:sous}

To study the morphology and kinematics of the CO cavity we are interested in the $^{12}$CO(2-1) arising from the outflow only. However, the $^{12}$CO(2-1) emission mixes emissions from both the disk and flow (see Fig.~\ref{f:12co}a). In the following, we describe the method we used to subtract the emission arising from the disk from the global $^{12}$CO(2-1) emission. 

We confirm at much higher sensitivity the result of P06 that the CO outflow triggered by HH30 is strongly monopolar (see Fig~\ref{f:12co}b), only clearly detected towards the north-east and we showed in the previous section that the $^{12}$CO emission arising from the south-western hemisphere is dominated by disk emission. In addition, the $^{12}$CO transverse pv diagrams shown in Fig.~\ref{f:12co-pvxy} are characteristic of a disk in Keplerian rotation for y $<$0 (e.g the panel at y=-0.3$^{\prime\prime}$), while the longitudinal pv diagrams in Fig.A.3 show symmetric contributions from the two faces of the disk at x $\pm$ 0.6$^{\prime\prime}$. Therefore we assume that the $^{12}$CO emission arising from the bottom hemisphere originates from the disk alone and that the disk emission is symmetric in $^{12}$CO with respect to the disk mid-plane as it is in $^{13}$CO. This is supported by Fig.~\ref{f:13co-vs-12co}, which clearly shows that the $^{12}$CO emission closely follows the $^{13}$CO emission in the south-western hemisphere. We then use the bottom $^{12}$CO emission to estimate and subtract the disk contribution in the top hemisphere.

\begin{figure*}[h!]
\centerline{
\includegraphics[width=1\textwidth]{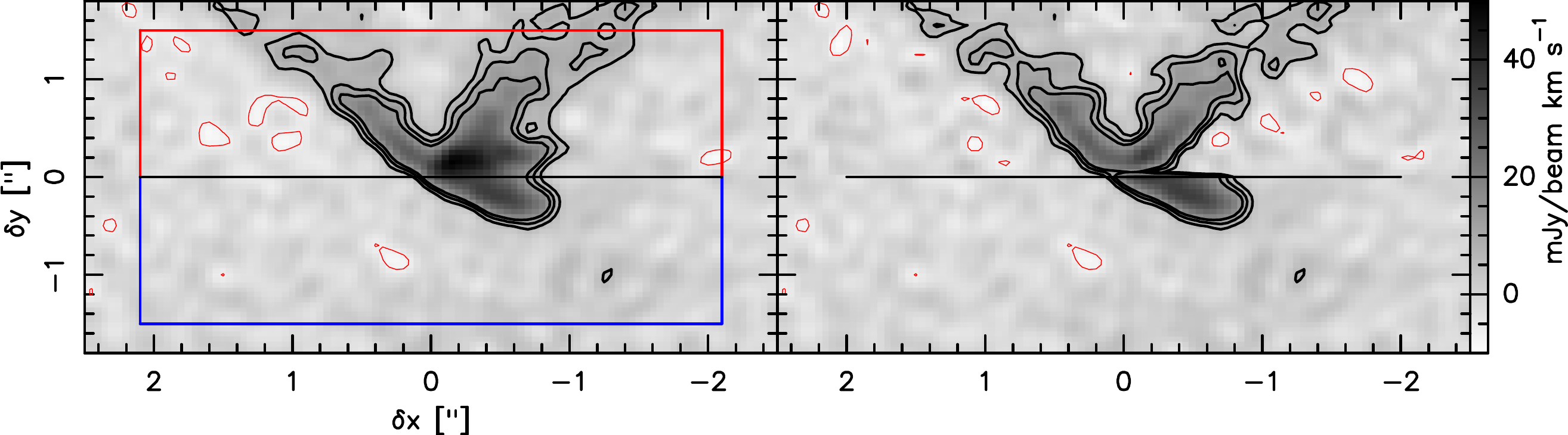}}
\caption{Illustration of the disk emission subtraction in the northern hemisphere. \textbf{Left:} $^{12}$CO emission of HH30 at 9.6 km s$^{-1}$ before subtraction of the disk contribution. The \textbf{right} panel presents the same channel after subtraction of the southern emission (blue rectangle on the left panel) from the northern emission (red rectangle on the left panel). The black contours show the +3$\sigma$, +5$\sigma,$ and +7$\sigma$ levels, while the red contours outline negative contours at -3$\sigma$, -5$\sigma$, and -7$\sigma$ with $\sigma$=2.0 mJy/beam km s$^{-1}$.}
\label{f:sousdisq}
\end{figure*}

Figure~\ref{f:sousdisq} illustrates for one particular channel (at 9.6 km s$^{-1}$) the procedure used and its result. The disk-subtracted channel maps are shown in Fig.~\ref{f:12cochansub}. The emission at a given vertical offset $-y$ is subtracted from its symmetric position at a positive $+y$ offset. This subtraction is performed pixel per pixel for all channels in a predefined rectangular area (illustrated in Fig.~\ref{f:sousdisq}). The bottom panels of Figure~\ref{f:12co} show the resulting total intensity maps at intermediate and high velocities.
The total $^{12}$CO(2-1) flux of the outflow integrated above the 5$\sigma$ level is of 11.9 Jy.km s$^{-1}$. The outflow accounts for more than 75\% of the total $^{12}$CO emission.

\subsection{Mass and temperature distribution}
\label{ss:massf}

In this section, we estimate the mass of the conical CO cavity from the disk-subtracted $^{12}$CO(2-1) emission (see Sect.~\ref{ss:sous}).
The non-detection of the outflow in $^{13}$CO(2-1) down to our 3$\sigma$ sensitivity implies a flux ratio $^{12}$CO/$^{13}$CO$>$70. Therefore, the $^{12}$CO(2-1) emission is optically thin. In these conditions, and neglecting the cosmic background contribution (justified as $T_{\rm mb}>T_{\rm BG} \sim$3K), the column density of CO molecules in the upper level (averaged over the observing beam) is given by 
\begin{equation}
N_{\rm up}=\frac{4\pi}{\rm hc}\frac{2k}{\lambda_{ij}^2}\frac{1}{A_{ij}}\int T_{\rm mb} \,dV
,\end{equation}
where $h$ is the Plank's constant, $k$ the Boltzmann's constant, $c$ the speed of light, $\lambda_{ij}$ the wavelength of the transition considered, A$_{ij}$ its Einstein coefficient, $T_{\rm mb}$ the main beam antenna temperature in K, and $V$ the radial velocity in cgs units. For the $^{12}$CO(2-1) transition, $\lambda_{21}=1.3$ mm and A$_{21}=6.910\times 10^{-7}$ s$^{-1}$. The mean $\int T_{\rm mb} dV$ in the area where the CO(2-1) integrated emission exceeds 5$\sigma$ (see thick grey contour in Fig.~\ref{f:12co}d) is 38.7 K~km~s$^{-1}$ which, reported in eq. (1) gives $N_{J=2}$=5.8$\times 10^{15}$ cm$^{-2}$. Making the assumption that the $^{12}$CO emission is governed by a single excitation temperature T$_{\rm ex}$, the population of rotational levels of CO follows a Boltzmann distribution and we have
\begin{equation}
N_{\rm CO} =\frac{N_{J=2}}{g_{J=2}  \times e^{-E_{J=2}/kT_{ex}}}\times \sum_{i=0}^{\infty}g_i e^{-E_i/kT_{\rm ex}}
,\end{equation}
where E$_i$ is the energy of level $i$, and g$_i$ the statistical weight of level $i$. The value N$_{\rm CO}$ has a minimum value 
at $T_{\rm ex}\sim$17K and then increases almost linearly with $T_{\rm ex}$ above 20~K owing the partition function. 
Since the emission in the cavity is optically thin, a lower limit to $T_{\rm ex}$ is given by the peak of the $^{12}$CO emission $T_{\rm mb} \simeq 30$K (after disk subtraction).
 Now, using $T_{\rm ex}$ = 30~K as lower limit for the excitation temperature over the outflow extension and assuming that the emission is governed by a single excitation temperature, we obtain a lower limit for the mean H$_2$ column density of $2.25\times 10^{20}$ cm$^{-2}$ with a CO abundance of 10$^{-4}$. On the given area of integration, $S=1.7\times 10^{-10}$ rad$^2$, at the distance $d=$140 pc of the source, it gives a minimal total mass for the CO cone of
\begin{equation}
M_{\rm CO} (30\,{\rm K}) =N_{H_2}\times S\times \mu\times m_{\rm H_2} = 1.7\times 10^{-5} \rm M_\odot 
,\end{equation} 
where $\mu$=1.4 is the mean gas weight per H$_2$ molecule, including 10\% of helium. 
This value compares well with the estimation of 2$\times 10^{-5}$ M$_\odot$ by P06 who integrated the emission of the outflow from -2$''$ up to +8$''$ (see their Fig. 6). 

\begin{figure}
\centerline{
\includegraphics[width=0.5\textwidth]{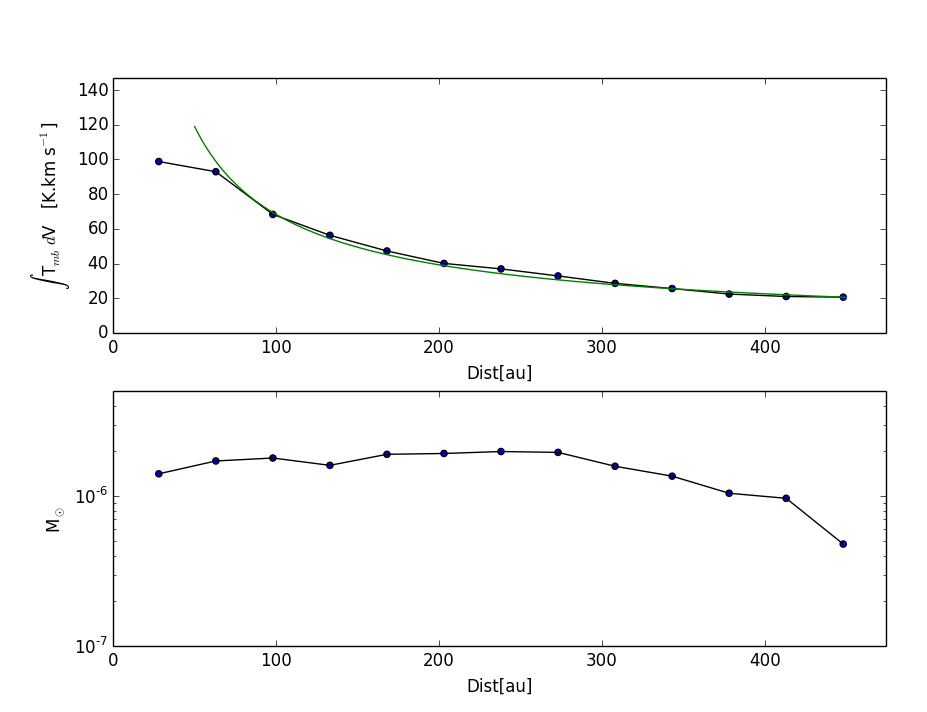}}
\caption{\textbf{Top:} Mean integrated line temperature of $^{12}$CO(2-1) in K km s$^{-1}$ (averaged over the area above 5$\sigma$ in each "slice" of Fig.~\ref{f:12co}) as a function of distance along the cone axis. The overplotted curve in green is defined by $T\propto z^{-4/5}$. \textbf{Bottom:} Mass in slices of thickness 30~au as a function of distance along the axis, using our lower limit to $T_{\rm ex}$ of 30~K. The constancy with $z$ could indicate a steady mass flux rate along the cone walls; see Sect.~\ref{ss:massf}.}
\label{f:massf}
\end{figure}

With the angular resolution given by ALMA, we have a total of 13 independent beams along the outflow axis. We may thus go further and investigate the variation of line brightness with distance from the source, by slicing the outflow extension into 13 independent measurements (see slices panel (d) of Fig.~\ref{f:12co}).
The velocity-integrated main beam temperature $\int{T_{\rm mb} dV}$ (line brightness), spatially averaged over each slice, is represented in the top panel of Fig.~\ref{f:massf}. It is seen to decrease exactly as $1/z^{4/5}$, where $z$ is the altitude above the disk.

On the bottom panel of Fig.~\ref{f:massf} we represent the mass in each slice of equal thickness $\Delta h$=30~au, obtained with $T_{\rm ex}$= 30~K and similar assumptions as for the derivation of total mass. Under these assumptions, the mass of the outflow per slice remains roughly constant as a function of distance. We attribute the drop beyond $2''$ (280 au) to the fact that beyond these distances one side of the cavity is becoming much fainter as seen in the channel maps of $^{12}$CO in Fig.~\ref{f:chan12-sous}.

\subsection{Modelling the transverse pv diagrams}
\label{ss:fit-ell}
 
 Our ALMA observations show  global morphological and kinematic properties similar to the PdBI observations of P06. In particular, both the channel maps and pv diagrams support the conical flow morphology derived by P06.  Using a global fitting procedure, P06 derived a half opening angle $\theta$=30$^{\circ}$~$\pm$~2$^{\circ}$, a constant outward flow velocity of $11.5~\pm~0.5$ ~km~s$^{-1}$,  and an inclination of -1$^{\circ}$~$\pm$~1$^{\circ}$ for the $^{12}$CO conical flow.  The P06 authors did not detect rotation but derived an upper limit of  Vrot~$<$ ~1~kms$^{-1}$ at radial distances from the jet axis of r=200~au. 

 Our higher sensitivity and angular resolution ALMA data allow us to search for rotation signatures that could have escaped detection in P06. Rotation induces a tilt in the pv diagrams transverse to the flow axis (see Fig.~15 in P06 and supplementary Fig.~1 in \citealt{hirota17}).  A small tilt is readily apparent in the ALMA pv diagrams shown in Figure~\ref{f:12co-pvxy} (e.g. at y=1.2$''$), in the sense that the blue-shifted emission peak (tracing the front side of  the cone) is slightly shifted spatially towards the south-east ($\delta x > 0$) with respect to the central flow axis position, while the red-shifted emission peak (tracing the back side of the cone) is shifted spatially towards the north-west ($\delta x < 0$). Such a tilt is consistent with flow rotation in the same sense as the underlying disk (see Fig.~15 of P06). We also search for wiggling signatures that would confirm the binary nature of the central source. 

\begin{figure*}
\centerline{
\includegraphics[trim=0cm 15cm 7cm 0cm, width=1\textwidth]{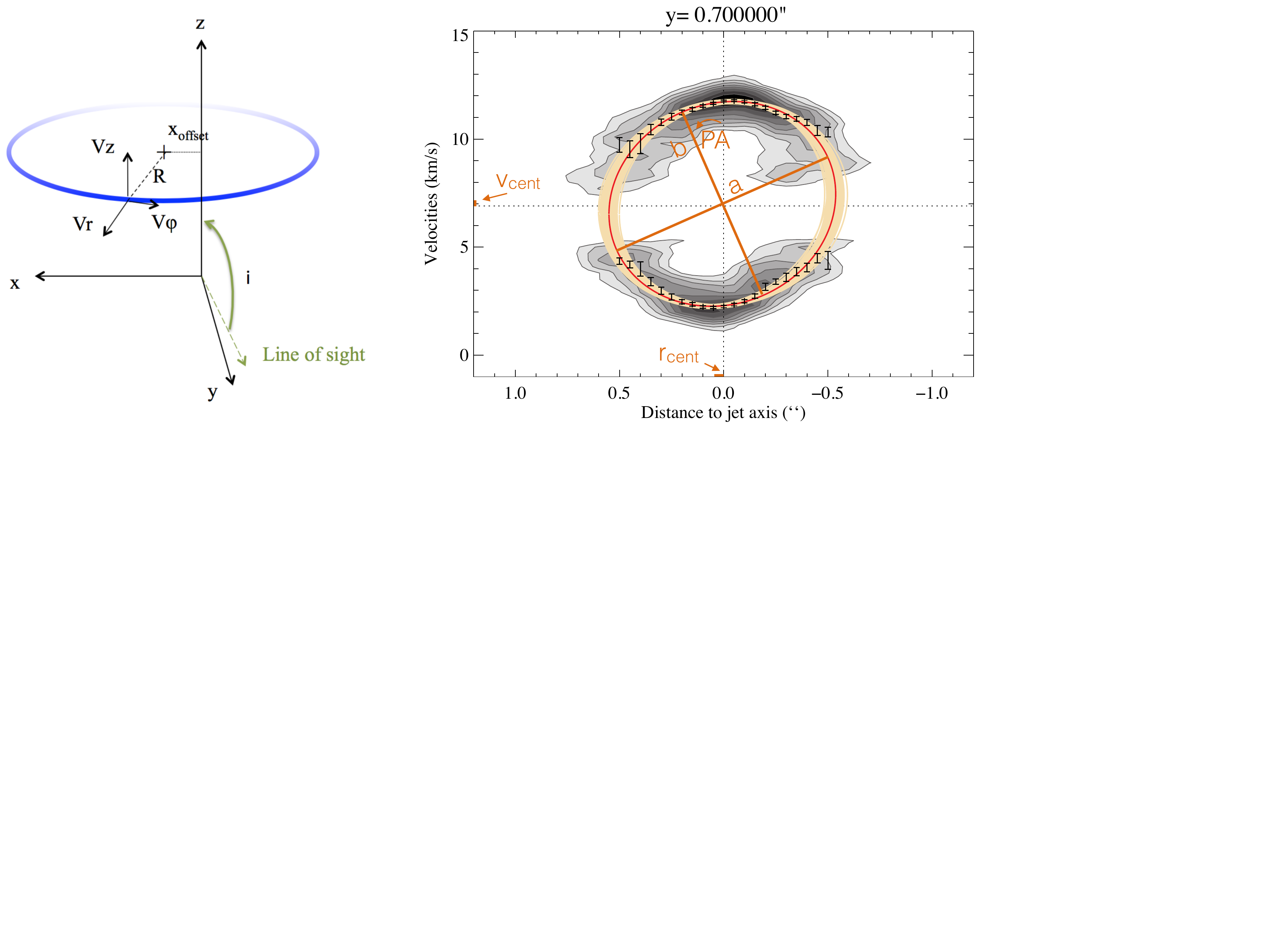}}
\caption{\textbf{Left:} Sketch of the ring model used to fit the transverse pv diagrams. The ring model is defined by five parameters illustrated in the figure and discussed in the text. \textbf{Right:} Illustration of the fitting procedure performed on the transverse pv diagram at z=+0.7$^{\prime\prime}$. The black crosses show the trace determined according to the method described in text. The red line shows the result of the fit of this trace by an ellipse. The yellow curves show the distribution of solutions taking into account the estimated 1$\sigma$ error on the five ring parameters. See text for more details.}
\label{fig:fitting}
\end{figure*}

  We take advantage of the very nearly edge-on orientation for the HH30 CO conical flow axis and fit the flow morphology and kinematics independently at each altitude z above the disk surface. This procedure is very similar to that recently used by \citet{hirota17} except that we allow for possible wiggling of the flow axis. This step is critical as not taking wiggling into account may induce spurious rotation signatures \citep{white14}. For each z, we assume that the $^{12}$CO emission arises from a narrow circular ring of gas defined by five parameters: radius R(z), centre x$_{\rm offset}$(z) , and velocity vector (V$_z$(z), V$_r$(z), V$_{\phi}$(z)) in the cylindrical coordinate system with axis z along the cone axis, seen at an inclination i to the line of sight. The parameter  x$_{\rm offset}$(z) measures the offset in the plane of the sky of the ring centre with respect to the flow axis. A sketch of the ring model is given in the left panel of Fig.~\ref{fig:fitting}. Such a ring projects onto a tilted ellipse in the transverse pv diagram. The ellipse is defined by five free parameters: semi-major and semi-minor axis, PA of the semi-major axis, and centre positions (in velocity and space).  The centre positions of the ellipse are direct measures of X$_{\rm cent}$(z) and V$_{\rm cent}$=$V_0+V_z~\times \cos({\rm i})$, where $V_0$ is the source velocity with respect to the adopted Vlsr=6.9 \kms ($V_0=0$ \kms in the absence of orbital motions). The full set of equations giving the transformation of the ellipse parameters into the ring parameters are given in Appendix~A. 
For each transverse pv diagram, the trace of the ellipse is determined through Gaussian fitting of the spectrum at each transverse distance $d$ from the axis. The blue-shifted and red-shifted wings of the profile are fitted separately. The range of velocities between 5 \kms and 10 \kms are excluded from the fitting because of possible contamination by residual disk/envelope emission. Error bars on the velocity centroids are estimated using the formula $\sigma$=FWHM/($2\sqrt{2\log{2}}\times$S/N), where FWHM is the full width at half maximum (in velocity) given by the Gaussian fit and S/N is the signal-to-noise ratio at the peak of the spectrum; the noise is estimated from the standard deviation of the adjacent continuum on both sides of the line. After careful inspection of the profiles, we used twice this theoretical uncertainty, which represented more accurately the error on the positioning of the Gaussian centre. An ellipse is then fitted to the trace using the IDL MPFITELLIPSE procedure. The determined ellipse parameters are then transformed into the shell radius, ring centre displacement from axis, and projected velocity components using the relationships given in Appendix~\ref{ap:ringfit}.

We derive the error bars on the final ring parameters (velocity components and offset position) in the following way. The IDL ellipse fitting routine provides uncertainties on the ellipse parameters taking into account the input error bars on the trace position.  We then compute 1000 sets of ellipse parameters by randomly drawing a realization for each parameter, assuming a Gaussian distribution with mean and standard deviation given, respectively, by the best-fit  solution and associated error bar provided by MPFITELLIPSE. We then transform these sets into 1000 sets of ring parameters using the relationships given in Appendix~\ref{ap:ringfit}.  We then derive the 1 $\sigma$ error bar for each ring parameter by taking the standard deviation of this resulting distribution. The results of this fitting procedure are illustrated on one selected pv diagram in Figure~\ref{fig:fitting}.

\subsection{Search for rotation signatures and wiggling}
\label{ss:res-fit-ell}

The fit is performed to each transverse pv diagram constructed every 0.15$^{\prime\prime}$ along the flow axis from z=0.1$^{\prime\prime}$ to z=1.75$^{\prime\prime}$. Values derived at z = 0.1$^{\prime\prime}$ above the disk plane are affected by large uncertainties because the ring radius is marginally resolved, hence the fit is not very good at this position. For distances beyond 1.8$^{\prime\prime}$, the front side (blue-shifted) of the conical outflow becomes significantly fainter than the red-shifted side and no satisfactory fit can be found. Reduced $\chi^2$ range between 0.6 and 2.3 except at the first position (z=0.1$^{\prime\prime}$ where it reaches 4). The fits to all transverse pv diagrams are presented in Appendix~A and Figs.~\ref{f:extra-fit-ellipse1} and \ref{f:extra-fit-ellipse2}.

\begin{figure*}
\centerline{
\includegraphics[width=1\textwidth]{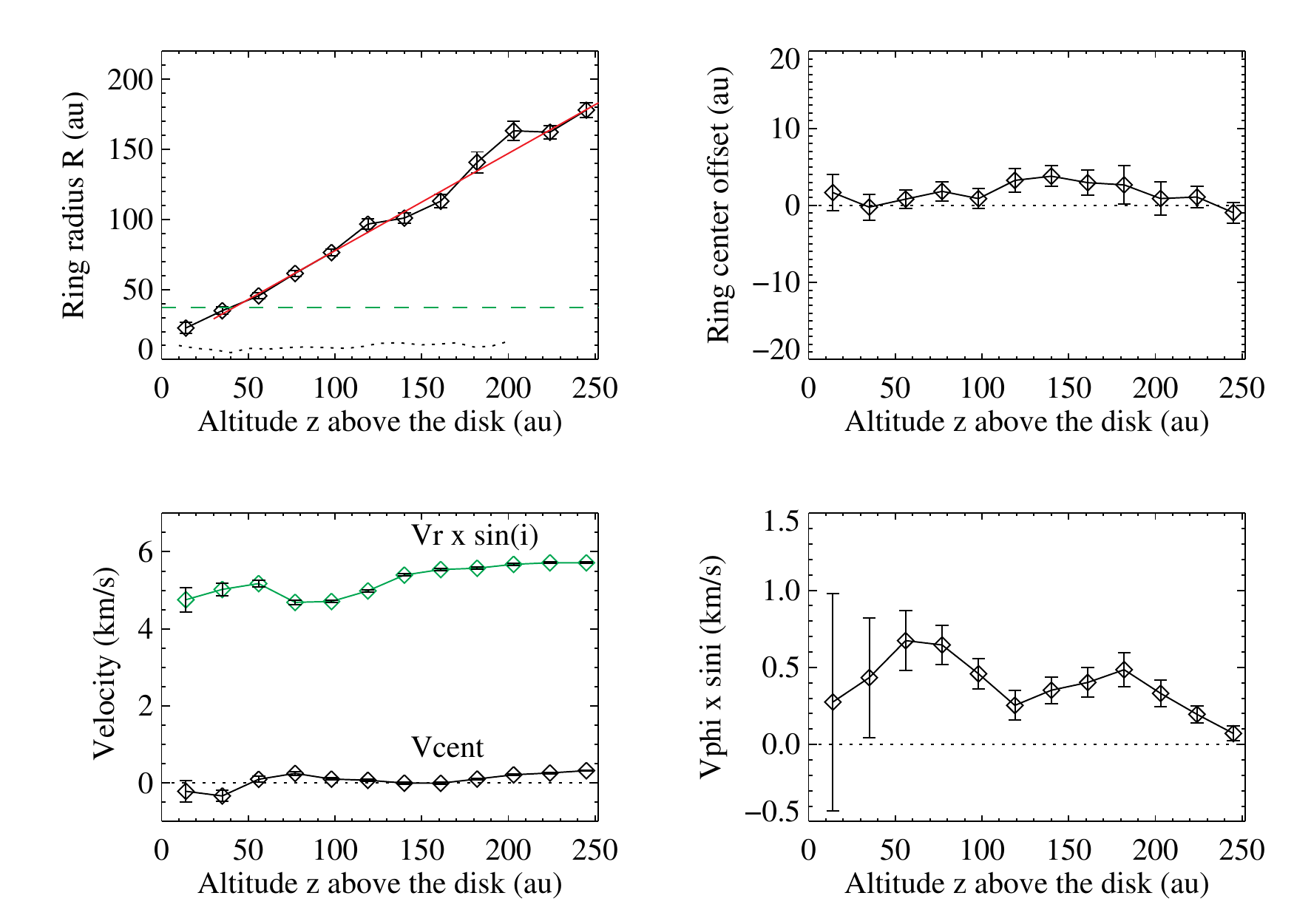}}
\caption{Variation of the ring parameters and their 1 $\sigma$ error bars derived from fitting the transverse pv diagrams, as a function of distance $z$ above the disk plane. \textbf{Top left:} Symbols show the CO ring radius as a function of $z$. The red line shows a linear fit giving a semi-opening angle of $35 \pm 1^{\circ}$. The radii increase linearly consistent with a conical geometry down to z=15 au.  The dashed green line locates the expected inner radius of the disk at $r$=37~au in the binary scenario from \cite{estalella12}. The dotted line plots the atomic jet radius derived from HST observations by \cite{hartigan07}. \textbf{Top right:} Variation of the ring centre displacement with respect to the continuum central $x$ position, as a function of $z$. \textbf{Bottom left:} Variation of V$_{\rm cent}$ (Black symbols) and  V$_{r}$ $\times$ $\sin{i}$ (Green symbols) as a function of z. Error bars include uncertainties on the source V$_{\rm lsr}$. Negative values of V$_{\rm cent}$ correspond to blue-shifted emission with respect to V$_{\rm lsr}$. \textbf{ Bottom right}: Variation of V$_{\phi}$ $\times$ $\sin{i}$ as a function of $z$. A consistent positive rotation signature in the same sense as the disk is detected in the central $\rm z \le 250~au$ of the flow.}
\label{fig:fitresults}
\end{figure*}

We plot in Figure~\ref{fig:fitresults} the five ring parameters derived from fitting the transverse pv diagrams and their variation with distance $z$ above the disk surface, assuming a distance of 140~pc for HH30.  The top panels show the evolution of the ring radius (left panel) and the ring centre displacement (right panel). The variation of radius is fully consistent with a cone of semi-opening angle 35$^{\circ}$ $\pm$ 0.7$^{\circ}$. The radius of the cavity keeps decreasing down to z=0.1$^{\prime\prime}$ confirming that the apparent cylindrical morphology of the emission in the channel maps below z=0.5$^{\prime\prime}$ is due to beam convolution effects. We derive an upper limit on the initial cone radius r$_{0}$ of the CO flow of 22~au, using the flow radius of 0.16$^{\prime\prime}$ measured at z=0.1$^{\prime\prime}$. We derive an average displacement for the centre of the ring in the x direction of 1.55~au over the central z $<$ 250~au. This is comparable to our estimated 1$\sigma$ uncertainty on the continuum image centring given by FWHM/(2$\sqrt{2\log{2}}\times$ SNR) = 0.75~au.  Some small amplitude wiggling may be present, however the displacements do not exceed 5~au and are compatible with zero within 3$\sigma$ at all but two positions.  We come back to this matter in the discussion section below.

 The bottom left panel plots the derived projected velocity components Vcent  and V$_r  \sin{\rm i}$. We use the usual convention for Vcent that negative values correspond to blue-shifted emission. The projected radial component of the velocity is well constrained and appears to be constant at V$_r$ $\times$ $\sin{\rm i}$ =  5.3 $\pm$ 0.4~km~s$^{-1}$ varying by less than 10~\% over the central 250~au. The derived values of Vcent are much lower, of the order of a few 0.1 km~s$^{-1}$. The mean value of Vcent=+0.07~km s$^{-1}\pm$0.09~\kms~over the central z=250~au is red-shifted but within our estimated 1$\sigma$ uncertainty on the source Vlsr (6.9~$\pm$~0.1~\kms).
  
\begin{figure}
\centerline{
\includegraphics[width=0.5\textwidth]{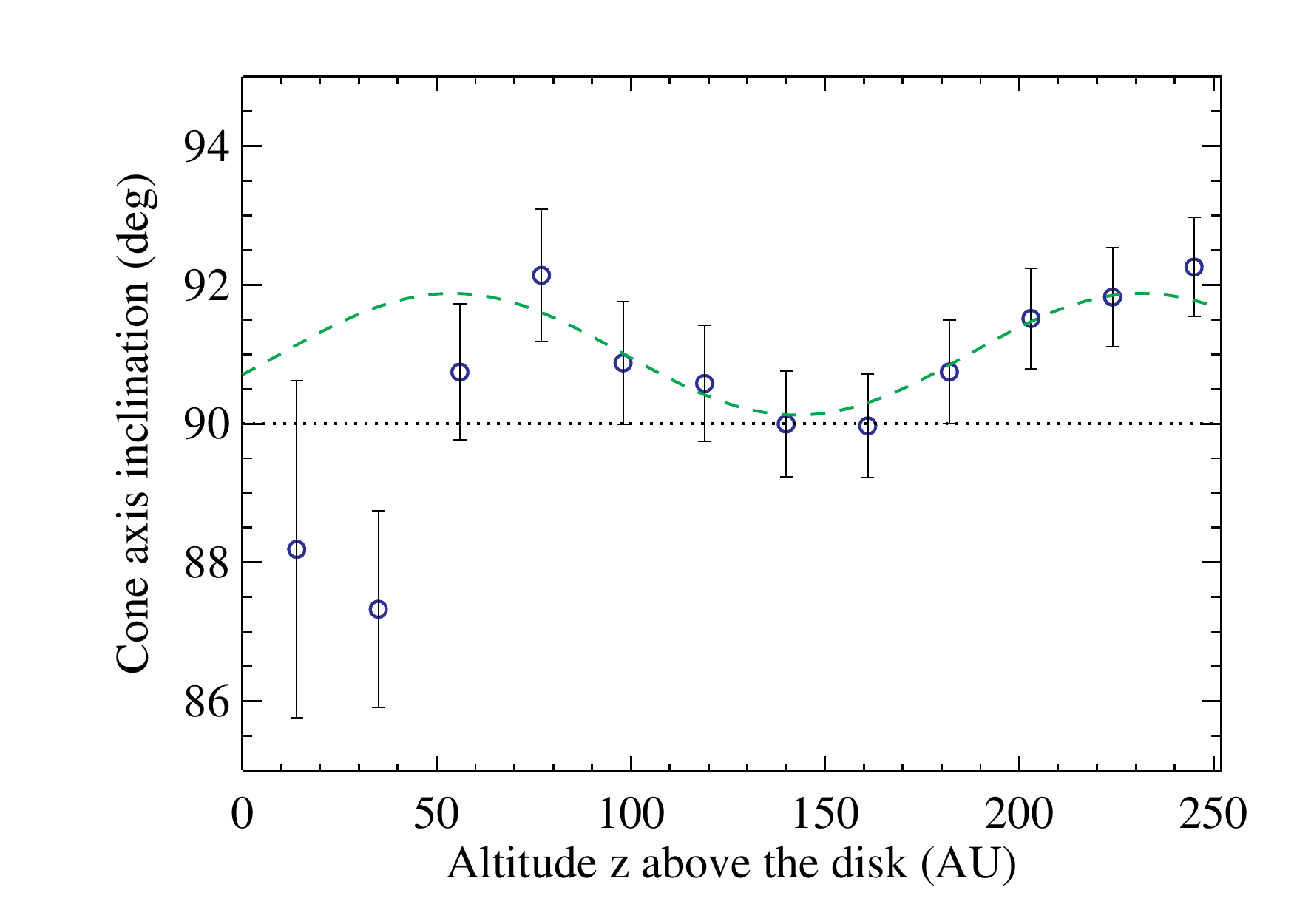}}
\caption{Variation of the CO flow axis inclination to the line-of-sight $i$ derived from the observed variation of $\frac{V_z \times \cos{\rm i}}{V_r \times \sin{\rm i}}$ assuming that the gas flows along the conical surface of semi-opening angle $\theta=35^{\circ}$. Plotted error bars include the 1$\sigma$ uncertainty on V$_{lsr}$ and $\theta$. The dashed green curve shows the prediction from the precession model that best fits the centre position and centroid velocity wiggling of the CO flow axis. See section~5.1.3 for more details.}
\label{fig:inclinaison}
\end{figure}

Under the assumption that the gas is flowing along the conical surface and that the orbital motions are negligible (V$_0 \simeq$0 \kms), we derive below the variation of the flow axis inclination implied by our observed variations of Vcent and V$_r$~$\times$~$\sin{\rm i}$. For a flow running along a cone of semi-opening angle $\theta$ the ratio of $Vr/Vz$ equals $\tan{\theta}$. If the cone axis is inclined at an angle $\rm i$ with respect to the line of sight, the ratio of the projected velocity components $\frac{V_r \times \sin{\rm i}}{Vcent}$= $\tan{\theta} \times \rm \tan{\rm i}$. We plot in Fig.~\ref{fig:inclinaison} the derived variation of the cone axis inclination with respect to the line of sight assuming an opening angle of $\theta=35^{\circ}$ for the conical surface as derived from Fig~12. The derived value of the CO axis inclination is well defined for projected distances z$\ge$~70~au and shows a remarkable sinusoidal variation around 91$^{\circ}$ with amplitude 1.2$^{\circ}$. The small amplitude variations of ${\rm i}$ observed along the flow can be understood as signatures of wiggling of the conical flow axis. We derive an average poloidal velocity ${\rm V_p}=\sqrt{{\rm V_z}^2+{\rm V_r}^2}$ = 9.3 $\pm$ 0.7 \kms. The small difference with the value of $11.5~\pm~0.5$ ~km~s$^{-1}$ previously obtained by P06 comes primarily from our larger semi-opening angle. 
  
Our derived mean $V_{\rm z} \cos{\rm i}$ value implies an average inclination of the CO flow axis with respect to the line of sight of 91$^\circ\pm1^\circ$ over the central z=250~au (including our 1 $\sigma$ uncertainties on V$_{lsr}$ of 0.1~\kms and on $\theta$ of 1$^{\circ}$), which is fully consistent with the previous derivation of P06.
   This is also consistent with the estimated inclination of the jet axis to the plane of the sky of $0\pm3^{\circ}$ inferred by \cite{burrows96}. However, \citet{coffey07} reported radial velocities of -5 \kms to -10 \kms for the northern atomic jet, suggesting inclinations of the jet axis to the line of sight of 84-87$^{\circ}$ for proper motions of the inner knots of 100~\kms estimated by \cite{estalella12}. Similarly, in the optical the upper side of the nebula always appears brighter than its lower counterpart \citep{stapelfeldt99}, suggesting that the northern disk axis is slightly tilted towards us. Therefore our finding  suggests a small tilt (a few degrees) between the CO flow axis and both the large-scale disk and atomic jet axis. However, within 3 $\sigma$ our ALMA observations are also compatible with the northern CO lobe being blue-shifted. We discuss the wiggling of the CO conical flow and its relation to the optical jet and the binary system in section 5.1.3.

\begin{figure}
\centerline{
\includegraphics[width=0.5\textwidth]{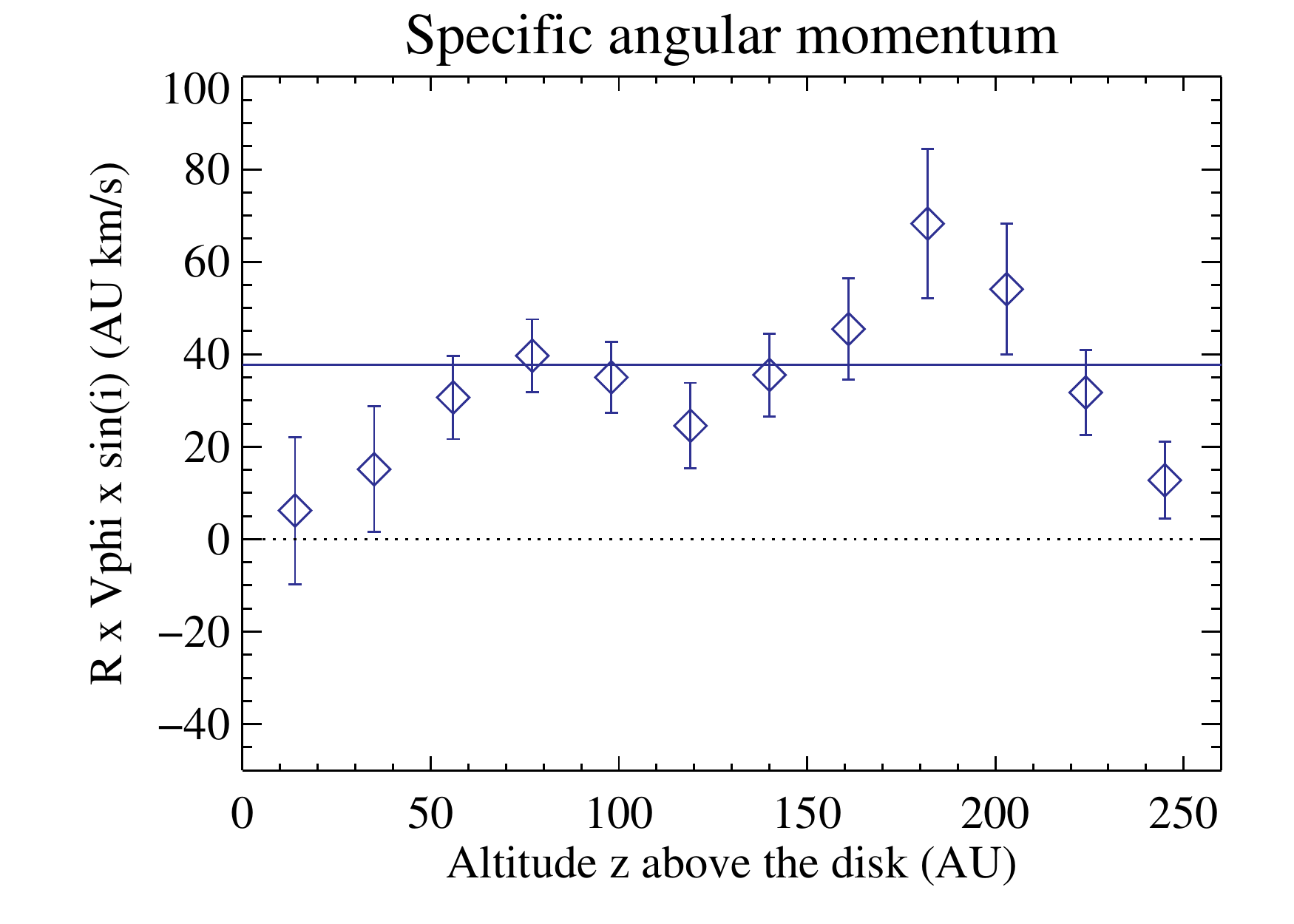}}
\caption{Variation of the specific angular momentum along the flow. Blue symbols show the variation along z of $R \times v_{\phi} \times \sin{i,}$ where R is the ring radius and V$_{\phi} \times \sin{i}$ the projected ring rotation velocity derived from the ellipse fits to the transverse pv diagrams. We derive an average specific angular momentum of +38~$\pm$~15~au~km s$^{-1}$ for ~50~au~$<$~z$<$~250~au.}
\label{fig:rvphi}
\end{figure}

Derived V$_{\phi}$ $\times$ $\sin{\rm i}$ are shown in the bottom right panel of Figure~\ref{fig:fitresults}.  A consistent positive rotation signature is detected along the flow in the central 250~au. The inferred sense of rotation of the CO cavity is consistent with the sense of rotation of the disk seen in $^{13}$CO. The derived  V$_{\phi}$ $\times$ $\sin{\rm i}$ are between 0.1 and 0.65 km~s$^{-1}$ and  drop as a function of distance from the source. We plot in Figure~\ref{fig:rvphi} the variation of specific angular momentum R $\times$ V$_{\phi}~\times ~\sin{\rm i}$ as a function of z, where $R$ is the derived ring radius. We derive an average specific angular momentum of 38~$\pm$~15~au~km~s$^{-1}$ over the central 250~au of the CO emission (excluding the first two points where V$_{\phi}$ is poorly constrained).

\subsection{Inner shell(s)}

For the inner shells visible at larger distances along the flow (y$\in[2'';5'']$), we do not attempt to perform detailed fits to the transverse pv diagrams because only partial emission along the ellipse is detected. At a few positions, it is however possible to fit an ellipse by hand to the transverse pv diagram. Such fits are just illustrative and constrain mostly the V$_r \sin({\rm i})$ and V$_{cent}$ components. The radius and V$_{\phi} \sin({\rm i})$ components remain poorly constrained.  The general trend observed is illustrated in Figure~\ref{f:pvxy22}. The inner shell systematically shows smaller $V_r \sin{\rm i}$ components and larger  V$_{cent}$ than the outer shell. Such behaviour implies a smaller opening angle $\theta$ for the poloidal vector of the inner shell with respect to the outer conical shell and could be expected from a more collimated inner shell of material. Red-shifted centroid velocities V$_{cent}$ are indicated for this inner shell. This could explain why the red-shifted part of the CO emission becomes stronger at distances $y \ge 2^{\prime\prime}$. A stronger displacement from the flow axis is also observed at some positions, suggesting a stronger wiggling for this inner shell. We discuss the possible relationship of these inner shells with the outer conical flow in the following section.

\section{Discussion: CO cavity and binary}
\label{s:discu}

Two broad classes of models have been proposed for the origin of the small-scale CO cavities at the base of molecular outflows: either the CO emission traces matter directly ejected from the disk (disk wind hypothesis) or entrained/shocked matter traces the interaction between an inner wind and an outer medium (either static or not).  We discuss below in detail the new constraints brought by our ALMA observations on these different scenarios. 
  
\subsection{Disk wind hypothesis}

As already shown by P06 our observed morphology and kinematics for the base of the CO outflow (z $\le$ 250~au) are fully compatible with gas flowing at constant velocity along a conical surface of a 
35$^{\circ}$ semi-opening angle. The derived poloidal velocity of 9.3 $\pm$ 0.7 km s$^{-1}$ significantly exceeds the velocity we would expect from infalling material ($\sim$2 km~s$^{-1}$ at r=200~au around 0.45~M$_{\odot}$ central mass). Therefore, under the assumption that the CO flows along the cone surface, the CO emission most likely traces the outflowing material that is directly ejected from the disk. The bottom panels of Fig.~\ref{f:massf} represent the mass of the outflow per slice of $\sim$30 au. If the outflow has a constant outward poloidal velocity of 9.3 km s$^{-1}$, hence a velocity along the z-axis of $v_z=7.5$ km s$^{-1}$, it corresponds to a crossing time through each slice $t_{\rm cross}$ = 30 au/$v_z$ = 20 yrs. The similar mass in each slice at $z<280$ au then translates into a roughly steady mass flux over the last 180 yrs of typical value $\dot{M}_{\rm CO} = \Delta M/t_{\rm cross}$\,$\sim 8.9 \times 10^{-8} M_\odot$ yr$^{-1}$. 

The material may be ejected either through purely thermal processes, i.e. photo-evaporating disk winds, or through a combination of thermal and magnetic processes. We discuss in turn these two models below. 

\subsubsection{Photo-evaporated disk wind (PDW)} 

One interesting hypothesis is that the HH30 CO flow traces a thermal photo-evaporated wind originating from the inner regions of the disk. The basic principles of disk photo-evaporation are the following. High-energy radiation  (UV and/or X-rays)  originating from the central accreting protostar heats the disk surface to high temperatures (10$^3$-10$^4$ ~K), well above the mid-plane temperatures. At sufficiently large radius  the thermal energy of the heated layer exceeds its gravitational binding energy and the heated gas escapes. The result is a pressure-driven flow, which is referred to as a photo-evaporative disk wind (PDW). Recently, PDWs have attracted considerable attention as they provide for a very efficient disk dispersal mechanism (see the recent PPVI review by \cite{alexander14} and references therein). 

The derived flow velocity ($\simeq$ 10 km~s$^{-1}$) and the conical morphology of the HH30 flow both match expectations from a thermal disk wind (see \citealt{font04}). Interestingly, in HH30 the accretion rate onto the central star, estimated from the mass flux measured in the central atomic jet assuming an average ejection to accretion rate ratio of 10 \%, is  $\simeq$ 2 $\times$ 10$^{-8}$ M$_{\odot}$ yr$^{-1}$ \citep{bacciotti99}; this is on the same order of magnitude as our lower limit estimate on the CO outflow mass loss rate of 9~$\times$~10$^{-8}$~M$_{\odot}$~yr$^{-1}$. Photo-evaporated disk winds are supposed to play an important role at the end of accretion processes when the disk accretion rate drops below the photo-evaporation rate. We may be directly witnessing this critical stage.

If the CO emission in HH30 traces a wind thermally ejected from the upper layers of the disk, we expect conservation of angular momentum along the streamline. Indeed, in such a wind there is no available torque to extract angular momentum from the disk. Thus, in the absence of strong turbulence, which could induce mixing between streamlines, the matter just carries away along its streamline the initial specific angular momentum inherited from its launching radius in the Keplerian disk. In a Keplerian disk the specific angular momentum increases with the disk radius as r~$\times$~V$_{\phi}$~ = ~30 au~km~s$^{-1}$ $\sqrt{(M_{\rm star} /1 M_{\odot} )~\times~(r/ 1~ \rm au)}$. Hence, under the assumption that the CO emission traces streamlines in a PDW, the specific angular momentum measured in the CO flow of $38 \pm 15$ au km~s$^{-1}$ implies a launching radius  r$_0$ in the range 1-7~au for M$_{\rm star}$ = 0.45 M$_{\odot}$.

In the case of an isothermal wind, PDW models show that the rate of mass loss per unit area $\Sigma$(R) peaks at the critical radius R$\rm _c$ $\simeq$ 0.1-0.2 $\times$ R$\rm _g$ \citep{font04}, where $\rm R_g= \rm GM_{\rm star}/\rm c_s^{2}$ is the (cylindrical) radius where the Keplerian orbital speed is equal to the sound speed of the hot gas. This critical radius typically ranges between 1 and 10 au for the various models investigated so far \citep{alexander14}. The total mass loss per radius interval peaks further away at r$\simeq$ 10~au for UV dominated models and at r $\simeq$ 30-40~au for X-ray dominated models. 

Putting aside the discrepancy of the radius of ejection of 1-7 au for HH30 versus 10-30 au in the models, the large mass loss rate estimate of 9~$\times$~10$^{-8}$ ~M$_{\odot}$ may be a challenge for such models. Integrated mass loss rates predicted by extreme UV dominated heating scale as \citep{font04}

\begin{equation}
\dot{M}_{w,EUV} \simeq 1.6 \times 10^{-10} \left(\frac{\Phi}{10^{41} s^{-1}}\right)^{1/2} \times \left(\frac{\rm M_{\star}}{1 \rm M_{\odot}}\right)^{1/2} {\rm M}_{\odot}{\rm~yr^{-1}}
,\end{equation}
where $\Phi$ is the flux of ionizing photons.
Inner disk holes of a few aus can increase the mass flux by an order of magnitude in the extreme ultraviolet (EUV) driven case \citep{alexander06}. Nevertheless, ionizing fluxes 3-4 orders of magnitude larger than typically assumed would be required in HH30 to account for the observed CO mass flux. This seems very unlikely. On the other hand, X-ray driven photo-evaporation rates are predicted to scale linearly with the central source X-ray luminosity \citep{owen11} as follows:

\begin{equation}
\dot{M}_{w,X} \simeq 6.3 \times 10^{-9} \left(\frac{L_X}{10^{30} erg s^{-1}}\right)^{1.14} \times \left(\frac{\rm M_{\star}}{1 \rm M_{\odot}}\right)^{-0.068}\rm M_{\odot}~yr^{-1}
.\end{equation}

An X-ray luminosity of $\ge$ 10$^{31}$~erg~s$^{-1}$ could therefore account for the mass flux derived for the HH30 CO cavity. HH30 remained undetected in the Chandra X-ray survey of Taurus conducted by \cite{gudel07}. However upper limits on its X-ray luminosity are difficult to derive because of the large uncertainty on the photoelectric absorption column due to its close to edge-on geometry. An X-ray luminosity of 10$^{31}$~erg~s$^{-1}$ necessary to explain for the mass loss of HH30 is on the upper end, but not incompatible with the L$_X$ distribution for accreting stars in Taurus.

Another difficulty would be to explain the survival of CO molecules in such a wind. Terminal speeds in hydrodynamical modelling of photo-evaporated winds reach 2-3 times the sound speed at the launching point \citep{font04}. To reach terminal velocity of 9 km~s$^{-1}$ would require sound speeds in excess of 3 km~s$^{-1}$ at the launching point, i.e. $\rm T_{gas} \geq 2000~K$ for molecular gas. Near-infrared emission lines of CO with temperatures of a few thousand K have indeed been detected in proto-planetary disks with emission radii extending at least up to 0.7~au \citep{najita03}. CO might survive at these temperatures if the wind dynamics and non-equilibrium chemistry are taken into account. Recently \cite{wang17} have conducted hydrodynamical simulations of photo-evaporative winds coupled with consistent thermo-chemistry and shown that indeed molecules such as CO can survive in the flow at relatively high wind temperatures owing to reactions that are out of equilibrium. The CO molecules could then cool down adiabatically very rapidly along the flow reaching temperatures of a few 10 K on the spatial scales probed by our ALMA observations (z=25-300 au). If we combine a velocity acceleration by a factor 3 with a radial expansion by a factor $>$ 10 (from r $\leq$ 4 au at the launching point to r $\ge$ 40 au), adiabatic cooling can provide  a drop in temperature by a factor $\ge$ 50 at z $\ge$ 50~au. Interestingly, on such spatial scales we observe peak $^{12}$CO brightness temperatures of 30~K. The non-detection of $^{13}$CO at these positions indicate optically thin emission, therefore the gas excitation temperatures may largely exceed 30~K on these spatial scales. Detailed observational predictions on the spatial scales probed by ALMA are required to test the photo-evaporated disk wind scenario fully. 

\subsubsection{Steady magneto-centrifugal wind}

Another possibility is that the wind is launched through magneto-centrifugal processes. In that scenario, a large-scale poloidal magnetic field threads the disk in the vertical direction.  Unlike PDWs, MHD outflows remove mass but also exert a torque on the disk surface, removing angular momentum from the disk \citep[e.g][]{pudritz07,alexander14}.  This mechanism has been proposed to account for the launching of the inner atomic jets but it could well extend to larger disk radii. Indeed, \cite{panoglou12} have shown that for accretion rates typical of the Class II phase, such a disk wind remains molecular for launching radii $r_0$ $\ge$ 1~au. Hence the CO outflow could trace the outer molecular streamlines in a radially extended MHD disk wind. This scenario has been recently suggested by \cite{hirota17} for the origin of the rotating SiO outflow of the massive young stellar object Orion source I. 

Under the steady assumption, \cite{ferreira97} and \cite{casse00b} computed the density and velocity structure of radially self-similar magneto-centrifugal disk winds.  The conical shape of the HH30 CO flow matches the expected shape of the streamlines before maximum radial expansion and recollimation towards the axis have been achieved (see Fig.~8 in \citealt{casse00b}). Substantial recollimation occurs at large $z/r_0$ , i.e. $\ge$ 100 typically, depending on the exact MHD solution, where r$_0$ is the anchoring radius of the magnetic surface on the disk. On the other hand, terminal poloidal velocities are reached much closer in, on spatial scales an order of magnitude smaller (see. Fig.8 in \citealt{casse00b}). Hence the base of an MHD wind essentially looks like a cone with constant outward radial velocity.

\begin{figure}
\centerline{
\subfloat{\includegraphics[trim = 5cm 0cm 0cm 0cm, width=0.4\textwidth]{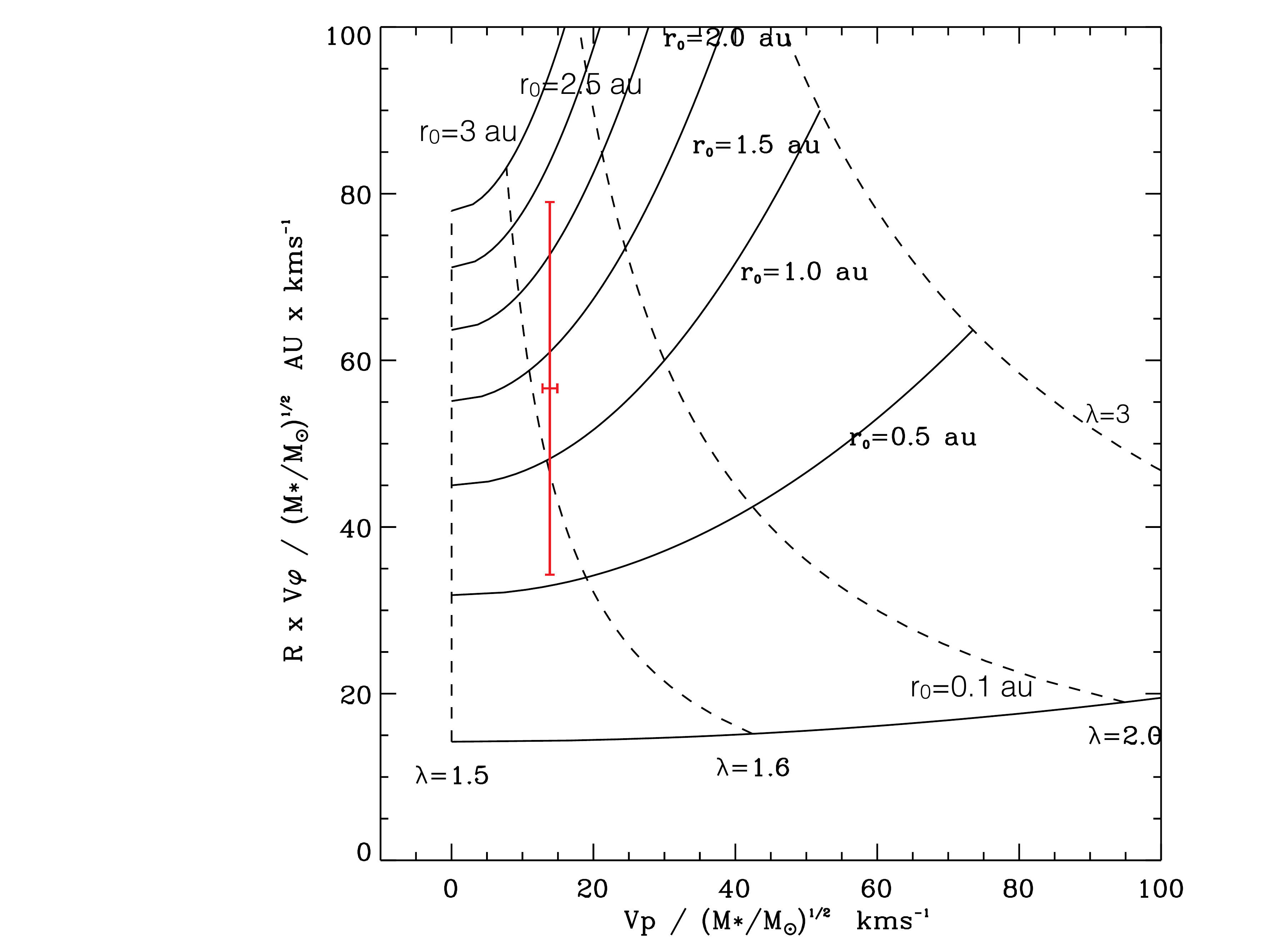}}}
\caption{Specific angular momentum r $\times$ v$_{\phi}$ vs. poloidal velocity V$_p$, both normalized to $\sqrt{M_*}$. Red symbols show the values derived for the HH30 CO cavity while curves show the expected relations from steady self-similar MHD disk winds launched from radii r$_0$=0.1,0.5,1,1.5,2,2.5,3~au (full curves) with magnetic lever arm $\lambda$=1.5,1.6,2.0, and 3.0 (dashed curves). Figure adapted from \cite{ferreira06}.}
\label{fig:fig_fdc06}
\end{figure}    

 In such a solution, the angular momentum is carried by both the field and the plasma.  At the disk surface, all the specific angular momentum is carried by the field, but it is completely transferred into the plasma afterwards. This transfer occurs rapidly after the Alfven surface on spatial scales z of  a few 10 $\times$ r$_0$. Afterwards the angular momentum of the matter is conserved along the streamline. In MHD disk winds that are steady, axisymmetric, and driven purely by magneto-centrifugal forces (negligible pressure) the asymptotic values of the flow specific angular momentum and poloidal velocity along a given magnetic surface are given by \cite{blandford82}
 \begin{eqnarray}
 r \times V_{\phi} & = & \lambda \sqrt{G M_* r_0} \\
 V_p & = & \sqrt{2 \lambda -3} \sqrt{G  M_*/r_0} 
 ,\end{eqnarray}
 where $r_0$ is the anchoring radius of the magnetic surface in the disk and  $\lambda$ the magnetic lever arm parameter of the solution (with $\lambda \simeq (\frac{r_A}{r_0})^2$ and $r_A$ the Alfven radius). To reach V$_p>0$ with pure magneto-centrifugal acceleration, lambda must be greater than 1.5.
 Taking  $r \times V_{\phi}$ = $38 ~\pm ~15$~au~km~s$^{-1}$ and $V_p= 9 \pm 1$ km~s$^{-1}$ we infer a launching radius of $r_0=$0.5-2.5~au and $\lambda$=1.6 (see Fig.~\ref{fig:fig_fdc06}). The smaller launching radii derived in the MHD disk wind scenario with respect from the PDW scenario derives from the fact that MHD disk winds always extract angular momentum from the underlying disk
(see Eq.~6 with $\lambda$~$\ge$~1.5). In PDW wind, the same equation applies with $\lambda$=1. Therefore, larger $r_0$ are required in PDW models to account for a given observed specific angular momentum. 
 
Combining Equ. (6) and (7) above, it may be seen that once a disk wind streamline has reached the asymptotic regime then the product $r\times V_\phi\times V_p$ is constant and only depends on  $\lambda$ and M$_*$ (see Eq.~10 in \citet{ferreira06}). Therefore if the atomic jet traces inner streamlines in the same MHD disk wind solution with $\lambda$=1.6, specific angular momentum less than 10  au~km~s$^{-1}$ for streamlines with poloidal velocities larger than 50~km s$^{-1}$ are predicted. Rotation velocities would be smaller than 0.7~km~s$^{-1}$ at radial distances from the jet axis r~$\ge$~14~au (angular resolution limit of HST observations). This would be consistent with the fact that no conclusive rotation signature was found for the collimated optical jet of HH30 with HST/STIS by \cite{coffey07} down to a precision of 5~km~s$^{-1}$.
 
 Such MHD disk wind solutions with small magnetic lever arm values are at the limit of the parameter space for steady magneto-centrifugal disk winds (for which the minimum authorized value of $\lambda = 3/2$). These solutions with small magnetic lever arm values correspond to slow and dense MHD disk wind solutions including a significant entropy deposition at the base of the wind \citep{casse00b}. Recent global non-ideal MHD simulations of 3D stratified disks have shown that such magneto-thermal disk winds may be a natural outcome of magnetized disks \citep{bai16,bethune17}.
 
In radially self-similar steady MHD accretion-ejection solutions where the disk wind extracts all of the angular momentum 
required for accretion, the radial variation of the mass accretion rate can be expressed with $\dot{M}_{acc} (r) = \dot{M}_{acc} (r_{in}) \times (\frac{r}{r_{in}})^\xi$ with $\xi \simeq \frac{1}{2 \times (\lambda-1)}$. The missing accretion mass flux that is extracted through the two-sided outflow, $\dot{M}_w$, is therefore given by

\begin{equation}
\dot{M}_{w} = \dot{M}_{acc}(r_{out}) - \dot{M}_{acc}(r_{in}) = \dot{M}_{acc}(r_{in}) \times \left[(\frac{R_{out}}{R_{in}})^{\xi} - 1\right]
.\end{equation}

We estimate the range of disk radii involved in the launching of the CO disk wind from the velocity width of the shell in the transverse pv diagrams. The FWHM in velocity of the profiles is typically 3~km s$^{-1}$, which translates into a range of poloidal velocities $V_p = 6-11$~km s$^{-1}$. Such a range of a factor 2 in poloidal velocity can be accounted for by a range of a factor 1/4 in disk radii (since $ V_p \propto V_{kep} (r_0) $). These shells could trace inner more collimated streamlines in the disk wind. With $\lambda = 1.6$ and $\frac{R_{out}}{R_{in}}=4$ we infer $\dot{M}_{acc}(r_{in}) \simeq 6 \times 10^{-8} M_{\odot} yr^{-1}$, comparable to the estimated mass flux in the CO wind. Such an estimate of the inner disk accretion rate would imply an ejection/accretion ratio of $\simeq 0.03$ for the inner atomic jet (estimated mass loss of $2 \times10^{-9}$~M$_{\odot}$~yr$^{-1}$; \citealt{bacciotti99}). This ratio is fully consistent with the typical ratio of jet mass-flux to accretion rate in T Tauri stars of 0.01-0.1 \citep{nisini18}. Therefore, it would suggest that the CO disk wind extracts most of the angular momentum flux needed for accretion across the wind launching region, and a large amount of the incoming mass-flux (since $\dot{M}_{w} \sim \dot{M}_{acc}(r_{\rm out})$). Detailed predictions in CO in such MHD disk wind solutions are required to confirm these order of magnitude estimates. 

With an inferred launching radius r$_0$ $\simeq$ 0.5-2.5~au, the wind is expected to be dusty, which helps the survival of molecules by shielding them from the protostar energetic radiation. \cite{panoglou12} computed the coupled ionization, chemical, and thermal evolution in a MHD disk wind solution with moderate magnetic lever arm ($\lambda$ = 14). These authors showed that for streamlines anchored at 1~au, the survival of CO molecules requires accretion rates above 10$^{-6}$ M$_{\odot}$ yr$^{-1}$. An MHD disk wind with $\lambda$ =1.6 as derived for the CO cavity in HH 30 would be significantly denser than the MHD wind solution investigated by \citet{panoglou12}, which could help to shield the CO molecules from photo-dissociating radiation.

An MHD disk wind origin for the low-velocity, V-shaped CO outflow has been suggested in a few recent studies. In particular rotation signatures have been reported in five cases so far; all of these were reported in younger Class I and Class 0 sources  \citep{launhardt09,zapata15,bjerkeli16,tabone17,hirota17}.
\cite{zapata15} and \cite{tabone17} recently conducted a detailed comparison with expectations from steady MHD disk winds. In particular, \cite{tabone17} showed that rotation signatures in the slow SO/SO$_2$ outflow in HH212 are best fitted by a MHD disk wind with a small lever arm ($\lambda$~$\leq$~5). \cite{zapata15} derived a similar low lambda value ($\lambda$=2) for the outflow around DG~Tau~B. These two cases therefore point towards a MHD disk wind solution similar to that inferred in HH30. Interestingly, the observed specific angular momentum and the derived launching radii for all these younger sources are all larger than those derived in HH30: r$_0$ ranges from $>$ 10~au in Orion source I \& DG Tau~B \citep{hirota17,zapata15}, 5-25~au in TMC1-A \citep{bjerkeli16}, and up to 40~au in HH212 \citep{tabone17}. This suggests a possible evolutionary scenario in the radial extent of the launching region of the wind. \cite{nolan17} recently developed semi-analytical models of the steady MHD disk wind launching regions incorporating all diffusion mechanisms in the disk. These authors show that properties of the disks strongly impact the radial extent of the wind launching region. In particular, increasing the disk surface density corresponds to launching regions at larger radii. Clearly, investigation of the disk-outflow connection in a larger sample is required to investigate further this aspect.

\subsubsection{Constraints on the HH30 binary in the disk wind scenario}

\begin{figure*}
\centerline{
\subfloat{\includegraphics[width=0.8\textwidth]{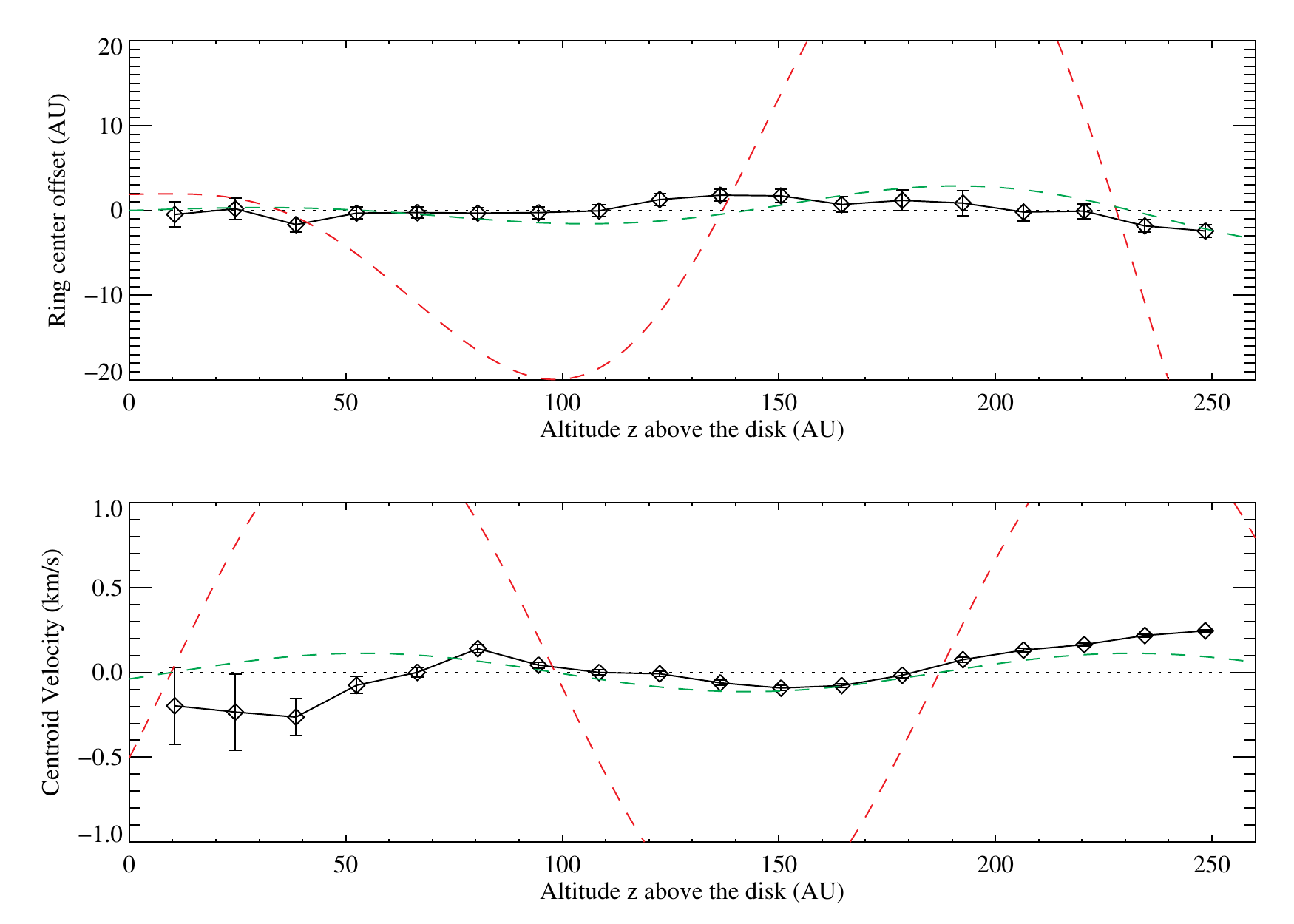}}}
\caption{Comparison of our derived CO flow axis wiggling (top panel) and centroid velocities variation (bottom panel) along the flow axis with the expectations of orbital (red) motions and precession (green) scenarios. Both measurements have been subtracted by their mean value (1.5~au for the transverse displacements and 0.07~km s$^{-1}$ for the centroid velocities). 
The red dashed curves shows the expected variations for the best fit orbital solution derived by \citet{estalella12}, assuming that the CO flow is arising from the same CS disk as the atomic jet. The green dashed curve shows the variations expected in the equivalent precessing model (see text for more details).}
\label{fig:sol_estalella}
\end{figure*}    

 Our ALMA observations bring new constraints on the binary scenario developed by \citet{anglada07} and \citet{estalella12} to account for the large-scale wiggling of the HH30 atomic jets. 
   
 \citet{estalella12} showed that the kinematics and positions of the knots in the HH30 atomic jet and counter-jet can be reproduced by the orbital motion of the jet source in a binary system with separation of $a = 18.0 \pm 0.6$ au whose orbital plane is perpendicular to the jet axis.  If the CO flow is indeed tracing a disk wind, we derive an upper limit for its launching radius of 7~au from its observed specific angular momentum. Therefore, in the orbital scenario favoured by \citet{estalella12}, the CO flow must be arising from one of the circumstellar (hereafter CS) disks. We plot in Figure~\ref{fig:sol_estalella} with a red dashed curve the expected flow axis wiggling and variation of the centroid velocity of the best-fit orbital solution derived by \citet{estalella12} at the time of our ALMA observations, assuming that the CO flow arises from the same CS disk as the atomic jet. In the orbital scenario, the flow ejection velocity and direction is assumed constant in time and the observed variation of $V_{cent}(z)$ is due to the orbital motion of the jet source. We take for the CO flow V$_{\rm z,CO}$=5.3 km~s$^{-1}$/tan(35$^{\circ}$)=7.5~km~s$^{-1}$ and for V$_{\rm jet}$=98~km~s$^{-1}$. We compare this prediction to the transverse displacements and centroid velocities of the CO rings as derived from our fitting procedure in the transverse pv diagrams. Both measurements have been subtracted by their mean value (1.5~au for the transverse displacements and 0.07~km s$^{-1}$ for the centroid velocities). The orbital solution clearly predicts too large wiggling both in position and velocities. Indeed the best-fit orbital velocity of the (primary) jet source has an amplitude of $1.5 \pm  0.2$~km~s$^{-1}$. If the CO flow was arising from the same CS disk as the atomic jet, we would expect the same variation of the CO profile centroid velocities, which is clearly not detected. This argument is also valid if the CO flow is arising from the CS disk of the companion (non-jet source) since its orbital velocity would be larger. Thus if the CO flow is indeed arising from a disk wind, it is not compatible with the 18~au binary orbital scenario favoured by \citet{estalella12}. 
 
We investigate below the equivalent precession solution that also reproduces large-scale wiggling of the atomic jet. In that scenario, the wiggling is produced by the precession of the flow axis, induced for example by the presence of a companion with an orbital plane inclined with respect to the disk plane of the jet source. This scenario was discarded by \citet{anglada07, estalella12} on the basis that it would require unrealistic small binary separations, $<$~1~au. We come back to this issue below. The precession model {\sl equivalent} to the previous orbital solution is defined by precession period$=$114~yrs, precession angle $\beta$ such that $tan(\beta)=v0/v_{jet}$, which gives $\beta=0.9^{\circ}$, precession phase $\phi_p$=$\phi_0+2*\pi$, and precession rotation sense opposite to the orbital rotation sense. Such a solution also reproduces the observed wiggling of the atomic jet axis positions on large scales. In the precession scenario, the flow ejection velocity is assumed constant in time and the observed variation of $V_{cent}(z)$ is due to the precession of the flow axis. We show in Figure~\ref{fig:sol_estalella} with a green dashed curve the expected flow axis position and centroid velocity variations for such a solution at the time of our ALMA observations, assuming constant V$_{\rm z,co}$=7.5~km~s$^{-1}$, V$_{\rm jet}$=98~km~s$^{-1}$. We also overplot in Figure~\ref{fig:inclinaison} the variation of the CO flow axis inclination predicted by such a solution assuming an average flow axis inclination to the line of sight of i=91$^{\circ}$. The precession solution strikingly reproduces the variation of the $^{12}$CO centroid velocity and derived flow axis inclination. This solution also reproduces the amplitude of variations for the flow  $x_{\rm cent}$ positions although there seems to be a small shift in phase between models and observations. The model predictions fall however within our estimated error bars on the $X_{\rm cent}$ positions. We note that to reproduce the line centroid velocities variation, a precession rotation in the sense opposite to the direction of rotation of the disk is required. Thus the CO flow axis wiggling and line centroid velocity variations are compatible with the same precession solution, which reproduces the wiggling of the atomic jet. This supports a scenario in which both the CO flow and atomic jet could arise from the same disk in solid body precession.

\cite{terquem99} investigated the scenario in which disk axis precession is induced by the presence of an outer companion with an orbital plane inclined with respect to the disk plane of the jet source. In that scenario, the misalignment between the two planes corresponds to the precession angle. Therefore, a small misalignment of only $\simeq~1^{\circ}$ would be needed to account for the wiggling of the CO flow axis. Equation~1 from \cite{terquem99} relates the precession period of the disk, assuming solid body precession, to the orbital period of the binary (see also Eq. 14 of \citealt{anglada07}). In that scenario the disk around the primary (identified as the jet/flow source) is truncated at $\simeq \frac{1}{3}$ of the binary separation. If we apply this equation and use the additional constraints on the total mass of the system ($M_{tot}$=0.45~M${_\odot}$) and assume that the wiggling due to the orbital motion of the CO flow source has to be negligible with respect to the precession motion of its disk axis (implying $V_0/V_{CO} \le \tan{\beta}$ hence $V_0~\le~$~0.2~\kms), we derive very small mass ratios between the secondary and the primary ($\mu \le 2 \times 10^{-3}$) and a binary separation ($\le$ 0.03~au). This solution appears very unlikely as the individual CS disk of the jet/CO flow source would be too small to launch either a jet or a massive CO outflow.

Alternatively, a precession scenario different from that studied by \citet{terquem99} could be operating. An inner non co-planar binary system could induce precession in the surrounding circumbinary (hereafter CB disk). Again, a small misalignment ($\simeq$~ 1$^{\circ}$) between the orbital plane of the binary and the CB disk plane would be enough to reproduce the observed precession angle. The atomic jet and the CO flow could then originate from different range of radii in this CB disk.  In that scenario, the inner binary would need to be very close to allow the launching of the fast atomic jet from the inner regions of the CB disk. 
Alternatively, the atomic jet could originate from the CS disk of one of the components in the close inner binary, while the CO flow would originate from the CB disk. This scenario was also suggested by \cite{tambovtseva08} from the non-detection of the large-scale CO wiggling in the PdBI observations of P06. The constraint on the launching radius of the CO outflow derived from our rotation measurements (r$_0$~$\le$~7 au) would imply a truncation radius for the CB disk $\leq$ 7 au. Hence the central binary would  need to be close (separation inferior to half truncation radius, typically) but such a separation allows the launching of a fast atomic jet from one of the CS disks. This scenario could also account for the small misalignment of a few degrees between the jet and CO flow axis. It is however not clear how such a configuration would lead to a very similar precession solution for both the atomic jet and CO flow. Another interesting precession mechanism was proposed by \cite{lai03}. Large-scale magnetic fields threading the accretion disk are required in MHD jet launching models. Such configurations may be subject to warping instability and a retrograde precession driven by the magnetic torques associated with the outflow. A detailed analysis in the context of the HH30 CO flow is required to test this mechanism.

\subsection{Entrainment scenario}
\label{ss:entrenement}

A second scenario for the origin of the flow, first discussed by \cite{pety06}, is that the CO conical structure in HH30 traces ambient gas swept up by a (so far unseen) wide-angle wind or by jet bow shocks. Such a scenario has been put forward to explain the V-shaped CO cavities commonly observed at the base of larger scale molecular outflows around much younger protostellar Class 0 sources \citep{ragacabrit93, lishu96, gueth99, lee01, arce07}. In the following, we discuss the constraints put on this scenario by our ALMA data.

We first note that our finding of a constant transverse velocity $V_r \simeq 5 $ km s$^{-1}$ in the HH30 outflow over a wide range in radius (from r=20 au out to r=100 au) is not consistent with the homologous expansion law $\overrightarrow{V} \propto \overrightarrow{Z}$ predicted by \cite{lishu96} for wide-angle wind driven cavities expanding into an ambient medium with density varying as $Z^{-2}$, where $Z$ is the spherical radius. This situation, later modelled numerically by \cite{lee01} and \cite{shang06}, clearly cannot describe the cavity kinematics in HH30. The very small implied cavity age $t_r = V_r / r \le$ 500~yr would also be a problem; it would imply that the wide-angle wind was launched only very recently compared to the source age of a few Myrs (class~II), which seems unlikely. 

However an ambient density distribution shallower than $1/R^2$ is expected after the onset of gravitational collapse on spatial scales z $\le$ 300~au especially at the age of HH30. \cite{delamarter00} studied numerically the interaction of an isotropic wide-angle wind with the flattened infalling and rotating envelope model from \cite{hartmann96}. These simulations predict the formation of conical shocked wind cavity. For the wind and ambient structure adopted in their simulations, an opening angle $\simeq 35^{\circ}$ as observed in HH30 is reached for $f^\prime=\dot{M}_{infall} / \dot{M}_{wind} = 10$. This value is in line with typical expectations that on average the disk accretion rate would verify $\dot{M}_{acc} \simeq \dot{M}_{infall}$ and $\dot{M}_{wind} \simeq 0.1\dot{M}_{acc}$. 

We estimated transverse expansion speeds predicted by the simulations of \cite{delamarter00} from the time snapshots of the dense wind case shown in their Figure~2. We derive typical cavity expansion velocities at $z \simeq 250$~au of 3 km s$^{-1}$ at 120~yrs and 1 km s$^{-1}$ at 160~yrs. Hence, in this shallower density gradient, the cavity expansion slows down rapidly over time and we expect that by the age of HH30 the shocked wind has reached pressure equilibrium against ambient gas, so the cavity reaches a steady state and no longer expands, while the shocked wind is forced to flow along its walls. The radial velocity $V_r$ is then not due to sideways expansion but to gas flowing parallel to the cavity walls. In this case the CO cavity age can be much larger than $V_r / r$. Such a steady-state wind cavity geometry was explored by \cite{barral81} for an isotropic wide-angle wind propagating in a self-gravitating disk atmosphere with vertical hydrostatic equilibrium. From their Figure~4, a final semi-opening angle of 35$^{\circ}$ from the pole would require $h / R = 0.5$, where $h$ is the altitude above the plane of the disk and $R$ the radius. Hence this stationary cavity scenario allows for longer ages, but requires material at large heights above the disk that is not detected from our $^{13}$CO and $^{12}$CO observations.

 The observed conical structure does not necessarily require a wide-angle wind. A stationary conical cavity could also be carved by a jet bow shock propagating into a $z^{-2}$ density field, as predicted analytically by \cite{ragacabrit93} and verified in the numerical simulations of \cite{cabrit97} (see their Fig.~4). Although no large-scale CO outflow is detected in HH30, we could only be sensitive to the very base of the brighter cavity. The detection of an inner shell of material on the distant (z $\ge$ 2$^{\prime\prime}$) transverse pv diagrams suggests the presence of an inner bow shock inside the main cavity, strongly supporting this possibility. The latter is also corroborated by the longitudinal pv diagrams at $-0.2''< x< 0.2''$, which  display a high-velocity component in both red-shifted and blue-shifted channels that could betray a recent bullet unresolved at our angular resolution. Finally, bow shocks are easier to confine since they are intrinsically more collimated and slower than wide angle winds. Stationary configurations are thus likely to be reached at lower ambient density with bow shock-driven cavities.  

 If the observed CO cavity mostly traces swept-up material, we can derive a crude estimate of the initial average density of the swept-up material by dividing the observed cavity mass by its volume (approximated by a cone of semi-opening angle of 35$^{\circ}$). We find n$_{H_2}$ $\simeq$ 3$\times$10$^{5}$~cm$^{-3}$. This value is comparable to the average density expected on 100~au spatial scales for a spherically collapsing envelope around a 0.5~M$\odot$ star with an infalling rate of 2.5 $\times$ 10$^{-7}$ M$_\odot$ yr$^{-1}$. Such infalling rate seems unlikely at the evolved stage of HH30. Moreover, with such densities, we would expect to detect a hint of the envelope in our $^{12}$CO and $^{13}$CO maps. Signatures of extended emission are detected in the channel maps close to the systemic velocity. However, this component appears to extend over the full ALMA field of view and does not seem peaked towards HH30. The recovered emission on scales of 10$^{\prime\prime}$ is weaker in $^{13}$CO than in $^{12}$CO, indicating it is less dense than 3~$\times$~10$^5$~cm$^{-3}$ (otherwise $^{13}$CO would have an optical depth of 200 at line centre for a line width of 1 km s$^{-1}$). Although ACA observations would be required to image this component fully and definitely rule out an envelope around HH30, it currently seems more likely that the CO cavity contains mostly ejected material.
 
High-angular resolution observations of the HH30 jet base conducted with HST by \citet{hartigan07} reveal the atomic jet morphology on scales very comparable to our ALMA CO observations. On the north-eastern jet side, proper motions are derived for three emission knots. Taking into account the 15 years time lag between the HST and ALMA observations, these three knots would lie at z=3.5$^{\prime\prime}$, 5$^{\prime\prime}$ and 6$^{\prime\prime}$ at the time of our ALMA observations. The  predicted positions of the first two knots overlaps with the positions where the inner CO shell is detected in our pv diagrams (z=2-5$^{\prime\prime}$). However a definite association between the radio and optical features is difficult. One reason for this is their different dynamical ages: a few years for the optical knots versus a few 100 years for the CO outflow.  New knots emerge every few years in the jet so it is likely that we have one or two new jet knots within the first 4$''$. Contemporary optical observations of the jet would be required to study in detail the connection between the atomic jet and CO outflow.

Other pending questions in the entrainment scenario are accounting for the observed precession and rotation of the CO outflow: Are they mostly inherited from the underlying atomic jet or from the surrounding medium (envelope/disk wind) in which the jet propagates? Detailed numerical simulations addressing these issues specifically are under way to fully test the entrainment hypothesis.

 \subsection{Origin of the asymmetry}
 
 As noted by \cite{pety06}, a peculiar aspect of the HH30 CO cavity compared to younger Class 0 outflows is the lack of detectable CO cavity on the southern side. We derive $^{12}$CO(2-1) flux ratios $>$ 10 between the northern and southern lobes of the CO cavity on spatial scales $\le$ 1$^{\prime\prime}$ implying an order of magnitude difference in column density.
 
 If the CO emission in HH30 is tracing swept-up material, it would imply a strong asymmetry either in the strength of the jet/wide-angle wind sweeping the cavity or in the ambient gas density on 40-200 au scales. \cite{bacciotti99} derived a similar mass and momentum rate between the HH30 jet and counter-jet, taking into account the difference in velocity recently derived by \cite{hartigan07} and \cite{estalella12}. Indeed, proper motions in the SW counter-jet are on average larger by a factor 1.7 to 2 than in the NE jet. The non-detection of the southern CO cavity is therefore  not likely due to a less powerful underlying jet. It also suggests that the faintness of the counter-jet is probably not due to larger extinction but maybe to less ambient material to entrain towards the South. This might also explain the asymmetry in velocities. The north-eastern atomic jet may be slower because the atomic emission partially traces entrained layers.
 
 The difference in jet/counter-jet velocity behaviour may provide an additional source of asymmetry for the CO cavity. The CO associated with the SW molecular cavity may be photo-dissociated owing to the faster and more variable SW atomic jet \citep{hartigan07,estalella12}. Indeed the receding jet shows significantly larger radial velocity variations, with $\Delta V \simeq$ 80-120~km s$^{-1}$ between knots, than the approaching jet where $\Delta V \simeq$ 50~km s$^{-1}$ \citep{estalella12}. Shock velocities larger than 100~km s$^{-1}$ can produce significant ionizing radiation leading to photo-dissociation of CO molecules in the SW molecular component while the NW molecular component would remain unaffected. However, it would be a bit surprising in that scenario to completely suppress the CO south-west cavity emission.
 
 The origin of the asymmetry could also be intrinsic to the launching process if the CO outflow traces a disk wind. Numerical simulations conducted by \cite{dyda15} explored the conditions under which asymmetric outflows/jets are created when combining rotating stellar magnetosphere with an inner viscous/diffusive magnetized disk. In particular, strong and persistent asymmetric disk winds arise in their simulations in the case of dense and weakly magnetized disks (with large $\beta$ representing the thermal pressure to magnetic pressure ratio in the disk mid-plane). However the inferred variability timescales are short (a few years) compared to the minimum timescale of the CO outflow in HH 30 ($>$ 500~yrs) and more in line with the timescales observed in the atomic jets. Similarly, recent MHD simulations  by \cite{bethune17} of magnetized disks performed on larger spatial scales and taking into account non-ideal effects show cases in which one-sided magneto-thermal disk winds can occur over specific range of disk radii. The understanding of such behaviour requires  more extensive numerical simulations exploring the full parameter space. In such models, the asymmetric launching results from a decoupling of physical conditions in the upper and lower surfaces of the disk.  No significant difference in the CO emission arising from each face of the disk is detected in our ALMA observations (see Fig.~\ref{f:13co}a). However, these observations explore the disk behaviour at radii 30 au $<$ r $<$ 250 au significantly larger than the spatial scales from which the flow would be ejected if it originates from a disk wind (r $\simeq$ a few au). Under this hypothesis, it remains to be explained why the CO outflow is monopolar while the atomic jet is bipolar. \cite{dyda15} found that the degree of symmetry of the outflows and their persistence both depend on $\beta$. A variation of this parameter with radius in the disk may lead to different properties of symmetry for jets/outflows launched from different radii. In addition the high-velocity jet could originate from the interaction of the  stellar magnetosphere with the inner disk, while the CO outflow would originate from much larger disk radii. In that scenario, different properties of asymmetry of the jet and outflow would be naturally expected depending on the nature of the star/disk interaction, and in particular, the location of the inner disk truncation radius, as well as the magnetization of the inner disk.
 
\section{Conclusions}
\label{s:concl}
We have observed the T Tauri pre-main sequence star HH30 with ALMA during the cycle 2 campaign in Band 6 (211-275 GHz). The circumstellar disk of HH30 is detected in continuum at 1.33 mm and in $^{13}$CO($J$ 2$\rightarrow$1), while the $^{12}$CO($J$ 2$\rightarrow$1) emission is a mixture of emissions arising from the disk and from the outflow.

The 1.3 mm continuum emission is fully resolved and shows an elongated morphology along PA=31.2$^{\circ}$~$\pm$~0.1$^{\circ}$ with a sharp fall-off in intensity at a radius of 75~au (0.55$^{\prime\prime}$). The emission is only marginally resolved in the transverse direction, implying an intrinsic vertical width $\leq$~24~au and an inclination to the line-of-sight ${\rm i}~\ge~85^{\circ}$. The continuum intensity profile along the disk is consistent with a constant flux of $2.4$ mJy/beam to within 3~$\sigma$. We do not detect an inner hole in the continuum image, unlike the previous finding by \cite{guilloteau08}. The $^{13}$CO emission line profile and $^{13}$CO channel maps are consistent a Keplerian disk, which agrees with the previous finding by P06. The $^{13}$CO emission is detected towards larger radii than the continuum emission, up to r=180 au. The upper and lower surfaces of the disk display very symmetric emissions with less than 15~\% discrepancy. From the $^{13}$CO integrated spectrum we derive a source $v_{\rm lsr}$ of 6.9~$\pm$~0.1~\kms.

The outflow of HH30 arises from the inner parts of the north-eastern surface of the disk and is detected in $^{12}$CO out to z=5$^{\prime\prime}$, or 700~au at the distance of the source. We derive a lower limit to the total mass for the CO outflow of $1.7\times 10^{-5}$ M$_\odot$. The channel maps and pv diagrams of the $^{12}$CO emission are consistent with the conical shell morphology previously derived by P06 out to z=1.8$^{\prime\prime}$=250~au. In addition, we detect signatures of an inner knot close to the source (z~$\simeq$~0.25$^{\prime\prime}$) and of an inner shell at large distances (z~$>$~2$^{\prime\prime}$).
      We confirm the conical shape of the cavity and derive a semi-opening angle of 35$^{\circ}$
      for 20~au~$<$~z~$<$~250~au. We constrain the base of the conical cavity at r$_0$~$<$~22~au. The derived velocity components are compatible with gas flowing along the conical surface with constant velocity V=9.3~\kms. We report detection of CO axis wiggling. The derived variation of the cone axis inclination to the line of sight shows a remarkable sinusoidal variation around 91$^{\circ}$ with amplitude 1.2$^{\circ}$ over the central z=250 au. We also detect small amplitude rotation signatures in the same sense as the underlying disk rotation sense with v$_{\phi}~\times \sin({\rm i})\in[0.1;0.7]$~\kms. We derive an average specific angular momentum  $r\times v_\phi = $38$\pm$15~au~km~s$^{-1}$ for  50~au~$< z <$~250~au.
      
    The morphology and the kinematics of the CO outflow are compatible with expectations from an origin in a slow disk wind, either through photo-evaporation or magneto-centrifugal processes. For both scenarios, we confirm the large minimum mass flux of 9$\times 10^{-8}$ M$_\odot$ yr$^{-1}$ for the CO wind.
      In the photo-evaporated disk wind scenario, conservation of angular momentum leads to a launching radius r$_0$ of 1-7~au, which is comparable to the estimated critical radii from which the mass flux starts to originate in these models. However, the derived large mass flux is difficult to account by current photo-evaporation models. On the other hand, an origin in a magneto-centrifugal disk wind implies a magnetic lever arm of 1.6 and launching radii in the range 0.5-2.5~au. Such MHD disk winds with small magnetic levers correspond to solutions including significant entropy deposition at the base of the wind. In both models, the wind extracts a significant amount of the accreted mass flux through the disk and likely plays an important role in the gaseous disk evolution.
      
     If the CO flow arises from a disk wind, our ALMA study brings new constraints on the central binary scenario in HH30. 
 The ALMA observations would rule out the orbital scenario previously favoured to account for the wiggling of the atomic jet, as it would predict centroid velocity variations of amplitude $1.5$~km~s$^{-1}$, much larger than observed. On the other hand, the equivalent precession scenario predicts centroid velocity and position variations much more in accordance with our ALMA observations. If the CO flow originates from one of the CS disks, unrealistically small separations of the binary are inferred. We therefore favour a precession scenario in which the CO flow originates from the CB disk around an inner non-coplanar binary with separation less than 3.5~au.
      
Another possible origin for the CO outflow is through entrainment of surrounding matter. If the CO cavity of HH30 is due to dragged material, the dichotomy between the age of HH30 (a few Myrs) and the constant radial velocity of the outflow of 5.3 km s$^{-1}$, which implies a cavity age $\sim$500 yr, favours a stationary cavity in which the material flows along the conical shape of the outflow. Detailed simulations for the evolution of the base of jet bow shock driven cavities on spatial scales comparable to our HH30 ALMA observations are under way to fully test this scenario and in particular to account for the observed rotation and wiggling of the CO outflow.

\begin{acknowledgements}
 \noindent{We thank the referee Francesca Bacciotti for a thorough report, which led to significant improvement of this paper. FL deeply thanks Edwige Chapillon for her vital help with the ALMA data set. FL thanks of A. Gusdorf for his help with the radiative transfer calculations. CD acknowledges Z.Y. Li for interesting discussions especially regarding the origin of asymmetry in the HH30 CO cavity. FL thanks Nicolas Cuello, Jorge Cuadra, and Daniel Price for constructive discussions about the binary of HH30. FL acknowledges support from the Joint Committee ESO government of Chile. FM and CP acknowledge funding from ANR of France (ANR-16-CE31-0013). CP acknowledges funding from the Australian Research Council via FT170100040 and DP180104235. }
 
 \noindent{This paper makes use of the following ALMA data: ADS/JAO.ALMA\#2013.1.01175.S. ALMA is a partnership of ESO (representing its member states), NSF (USA) and NINS (Japan), together with NRC (Canada), NSC and ASIAA (Taiwan), and KASI (Republic of Korea), in cooperation with the Republic of Chile. The Joint ALMA Observatory is operated by ESO, AUI/NRAO and NAOJ.}

\end{acknowledgements}

%
%

\bibliographystyle{aa}
\bibliography{fab}

\begin{appendix} 

\section{Complementary figures}

\begin{figure*}[h!]
\centerline{
\includegraphics[width=0.95\textwidth]{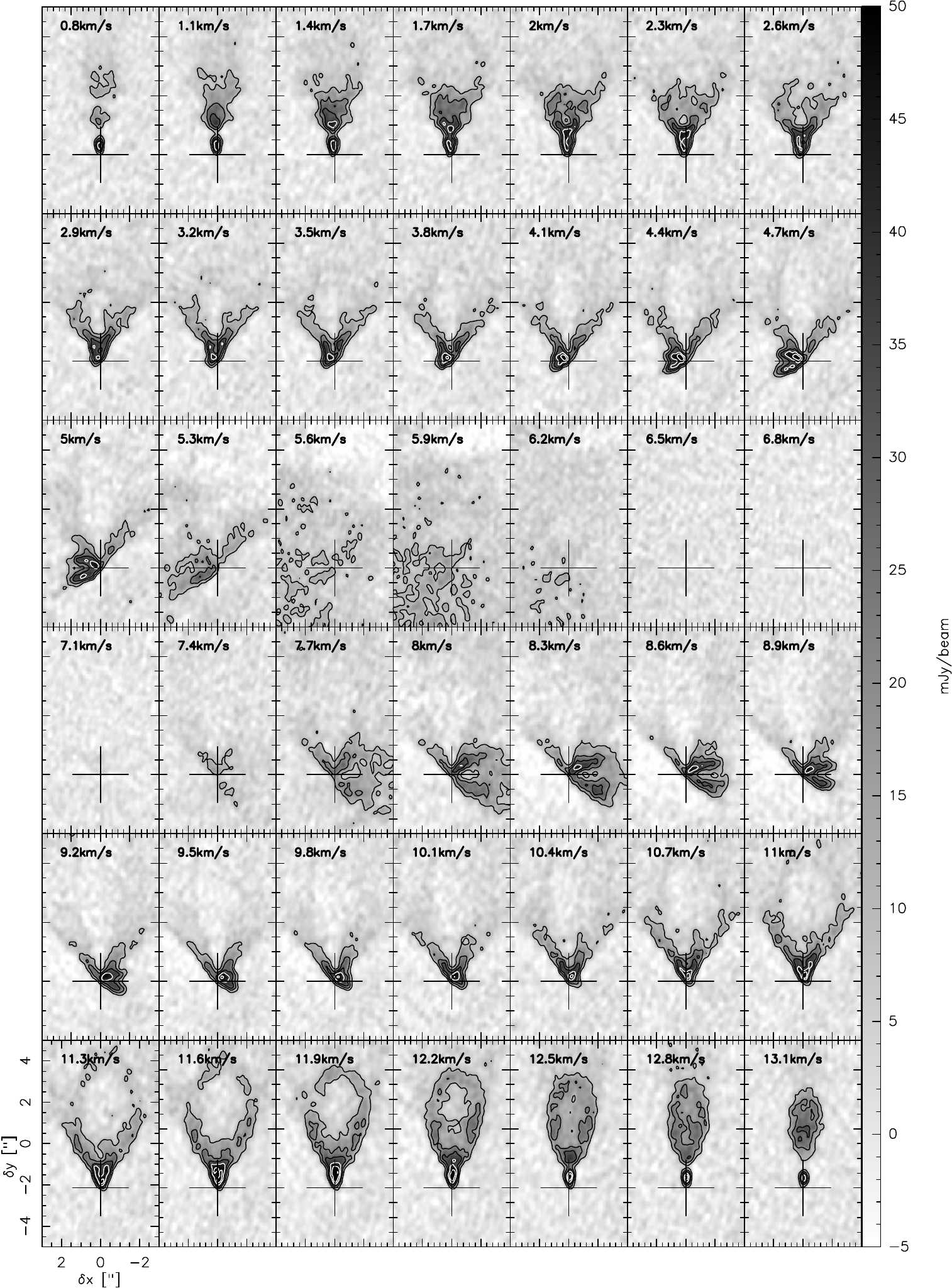}}
\caption{Channel maps of the $^{12}$CO(2-1) emission line of HH30. The contours start at 5$\sigma$ with 5$\sigma$ steps with $\sigma=2.0$ mJy beam$^{-1}$. The channel velocity is indicated in the top left corner in km~s$^{-1}$. The cross locates the central position of the disk.}
\label{f:chan12}
\end{figure*}

\begin{figure*}
\centerline{
\subfloat{\includegraphics[trim = 0cm 0cm 0cm 0cm , width=1\textwidth]{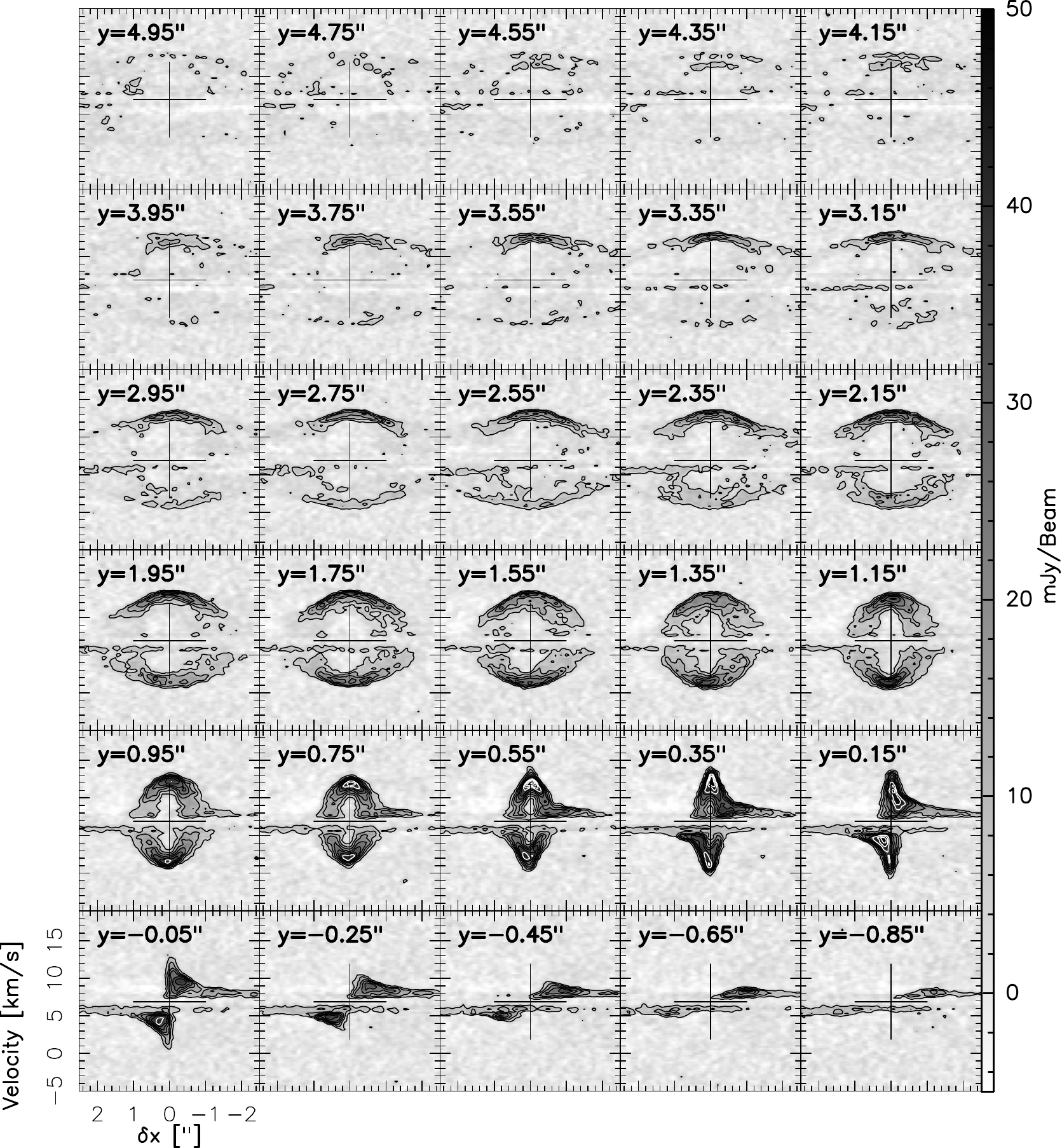}}}
\caption{Transverse pv diagrams of the $^{12}$CO emission, with pseudo slit parallel to the disk main axis, from y=+4.95$^{''}$ on the top left to y=-0.85$^{''}$ on the bottom right. The horizontal lines represent the $v_{\rm lsr}$ of HH30 at 6.9 km s$^{-1}$ while the vertical lines outline the position $x=0$. The contour levels start at 3$\sigma$ with 3$\sigma$ steps, with $\sigma=$1.86 mJy/beam.}
\label{fa:12co-pvxy}
\end{figure*}

\begin{figure*}
\centerline{
\subfloat{\includegraphics[trim = 1cm 1cm 12cm 0cm, width=1\textwidth]{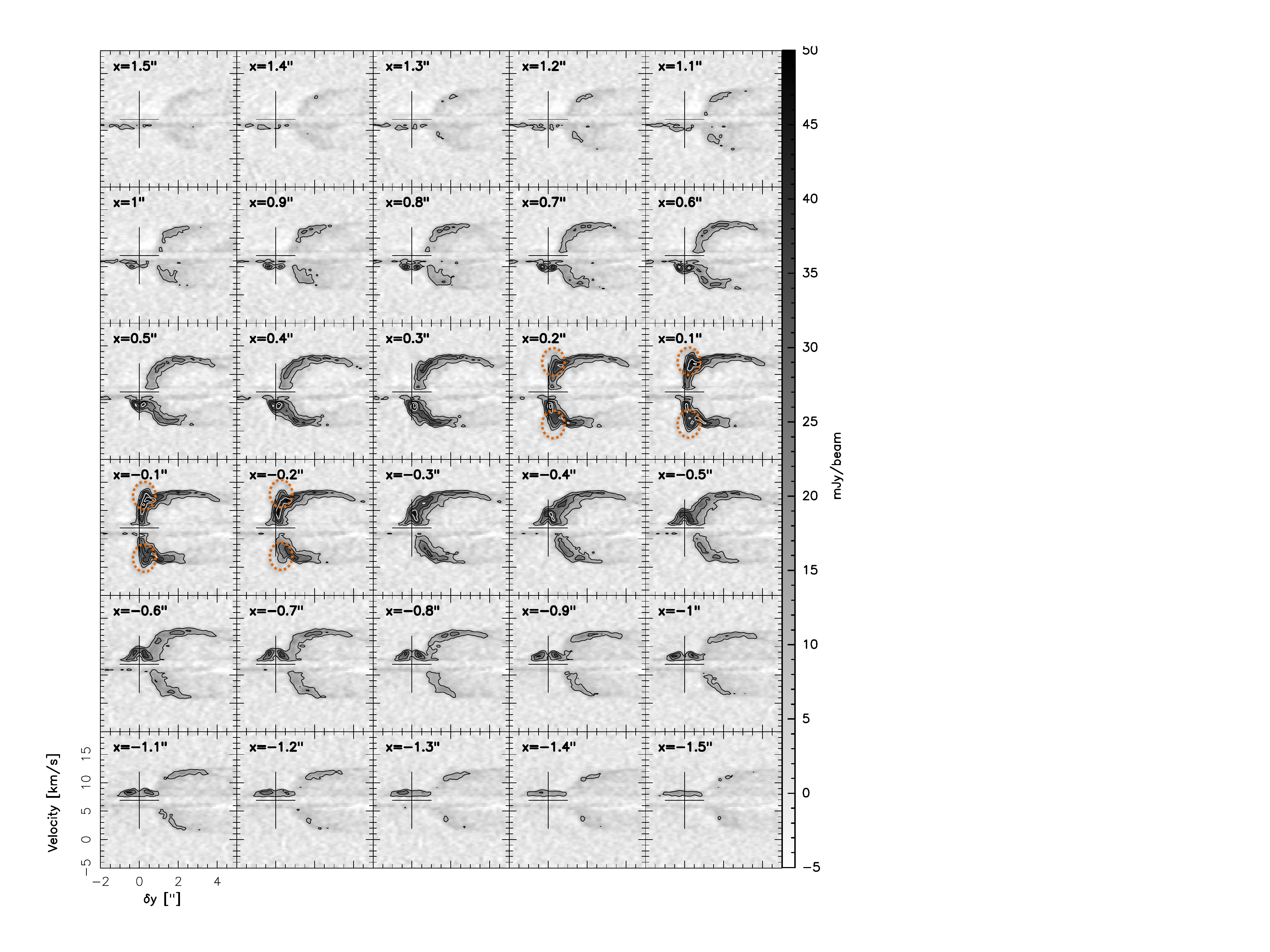}}}
\caption{Longitudinal pv diagrams perpendicular to the disk, starting at x=1.5$^{''}$ (top left) to x=-1.5$^{''}$ (bottom right). The black cross shows the central position of the disk at $x=0''$ and the $v_{\rm lsr}$ of HH30 at 6.9 km s$^{-1}$. The contour levels start at 5$\sigma$ with 5$\sigma$ steps, with $\sigma=$2.0 mJy/beam. The orange ellipse in dashed line highlights the high-velocity component seen at y$\sim$0.25'' from x=-0.2'' to x=+0.2''. See text for more details.}
\label{fa:12co-pvyx}
\end{figure*}

\begin{figure*}
\centerline{
\subfloat{\includegraphics[trim = 0cm 0cm 0cm 0cm, width=1\textwidth]{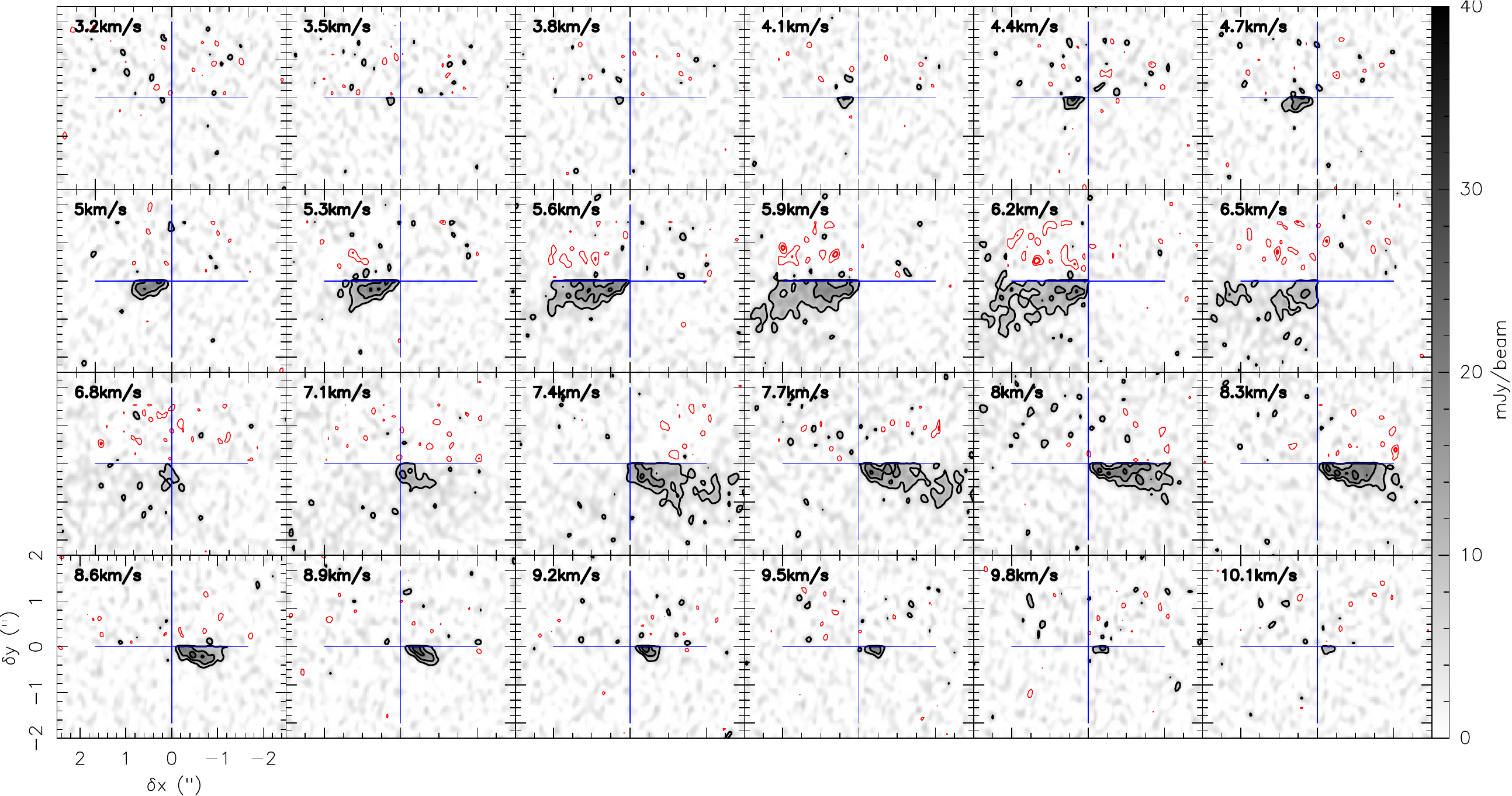}}}
\caption{Channel maps of the residual $^{13}$CO(2-1) emission line in the top disk hemisphere after subtraction of the symmetric emission from the bottom hemisphere. The contours in black start at 3$\sigma$ with 3$\sigma$ steps with $\sigma=2.3$ mJy beam$^{-1}$. The red contours highlight the negative emission at -3$\sigma$ and -6$\sigma$.}
\label{f:13co-res}
\end{figure*}

\begin{figure*}
\centerline{
\subfloat{\includegraphics[trim = 0cm 0cm 0cm 0cm, width=1\textwidth]{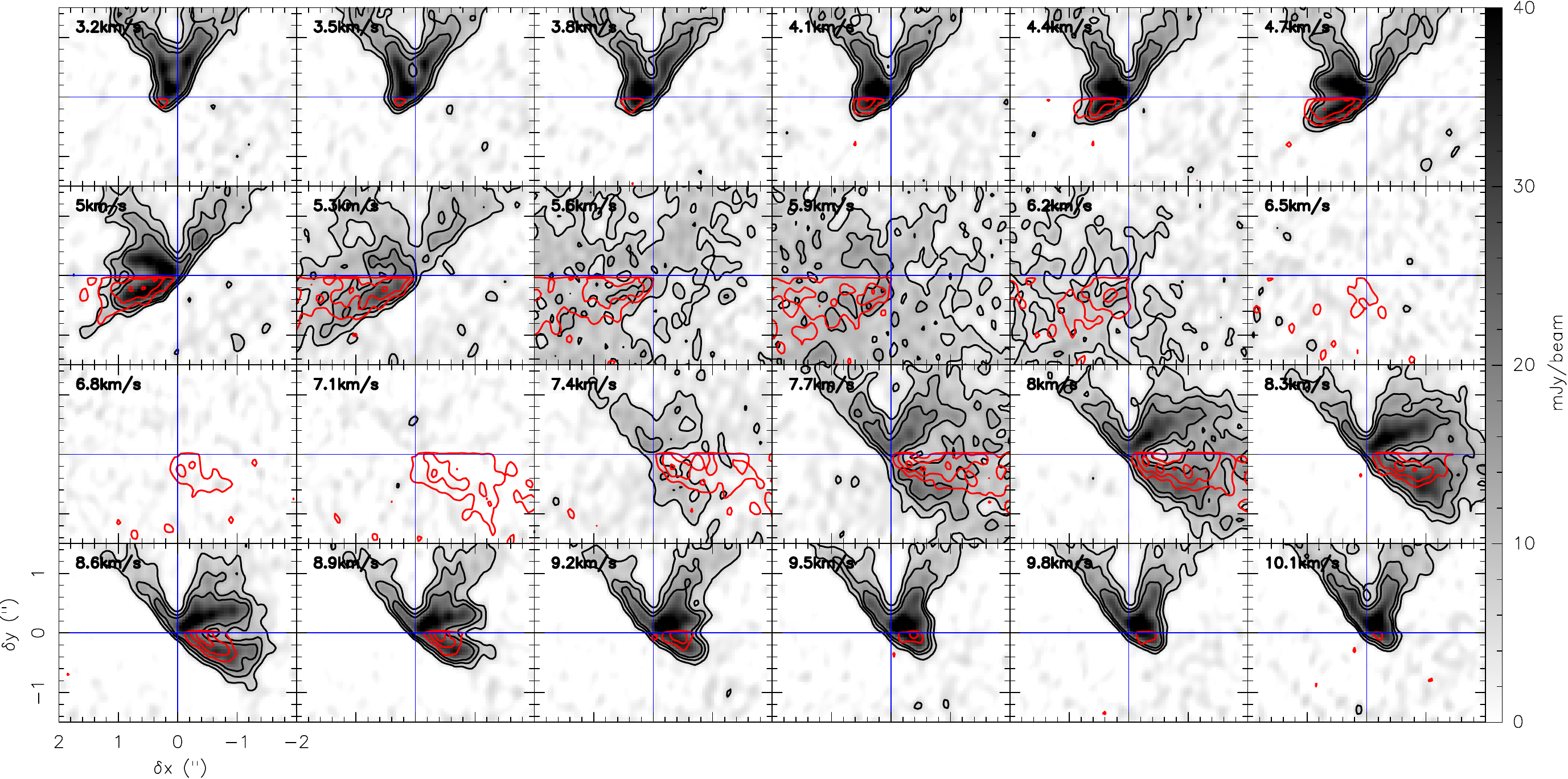}}}
\caption{Channel maps of the $^{12}$CO emission line of the disk of HH30 from 3.2 km s$^{-1}$ (top left) to 10.1 km s$^{-1}$ (bottom right). The black contours start at 5$\sigma$ with 5$\sigma$ steps, where $\sigma=2.0$ mJy Beam$^{-1}$. Overlaid in red are the contours of the $^{13}$CO(2-1) emission arising from the south-western part of the disk. The contours start at 3$\sigma$ with 3$\sigma$ steps, where $\sigma$=2.3 mJy beam$^{-1}$}
\label{f:13co-vs-12co}
\end{figure*}

\begin{figure*}[h!]
\centerline{
\includegraphics[width=0.95\textwidth]{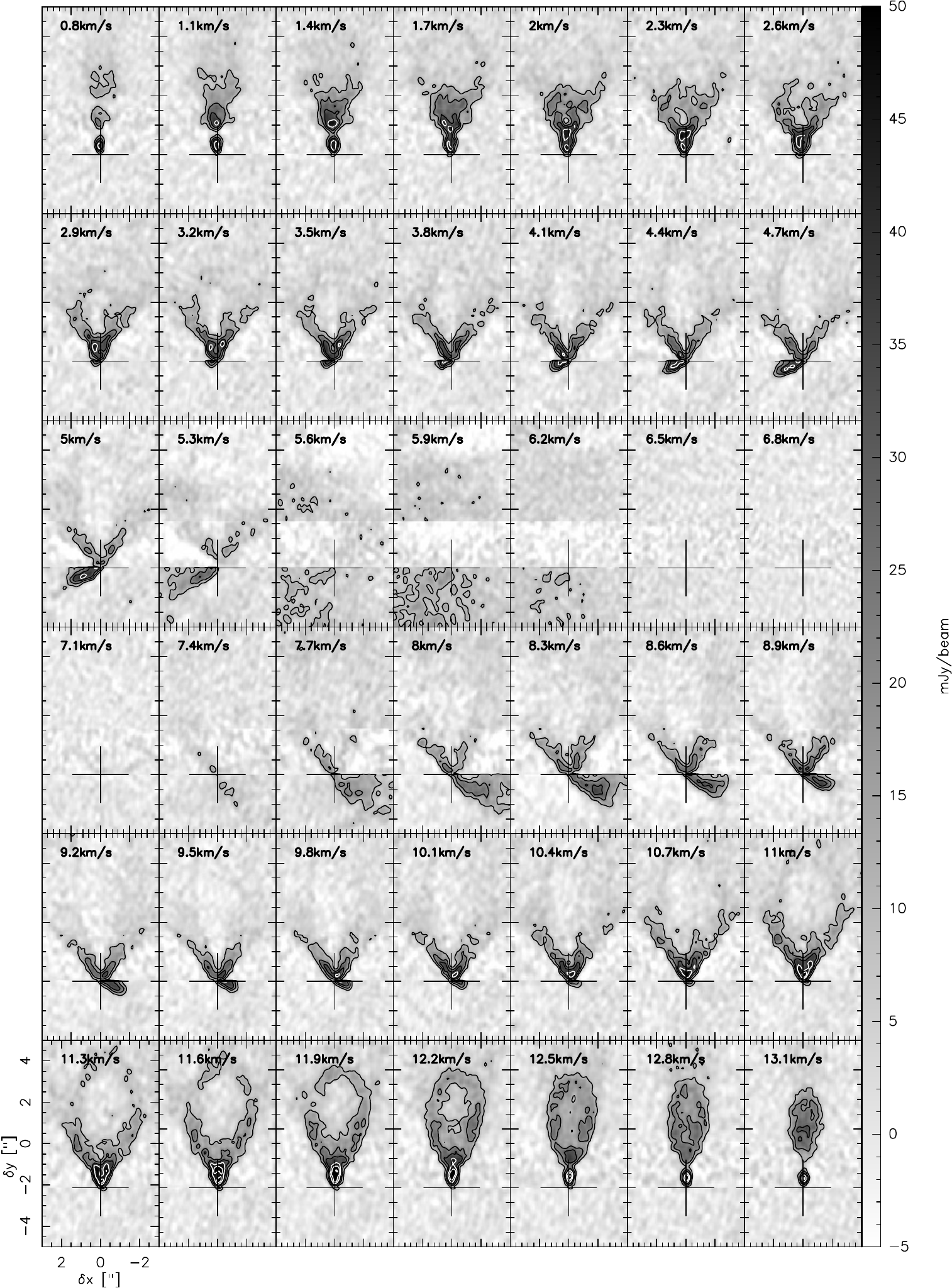}}
\caption{Channel maps of the $^{12}$CO emission line of the disk of HH30 after subtraction of the disk contribution in the northern hemisphere (see Sect~\ref{ss:sous}). The contours start at 5$\sigma$ with 5$\sigma$ steps with $\sigma=2.0$ mJy beam$^{-1}$ (or 0.8 K). The channel velocity is indicated in the top left corner in km~s$^{-1}$.}
\label{f:12cochansub}
\end{figure*}

\begin{figure*}
\centerline{
\subfloat{\includegraphics[width=1\textwidth]{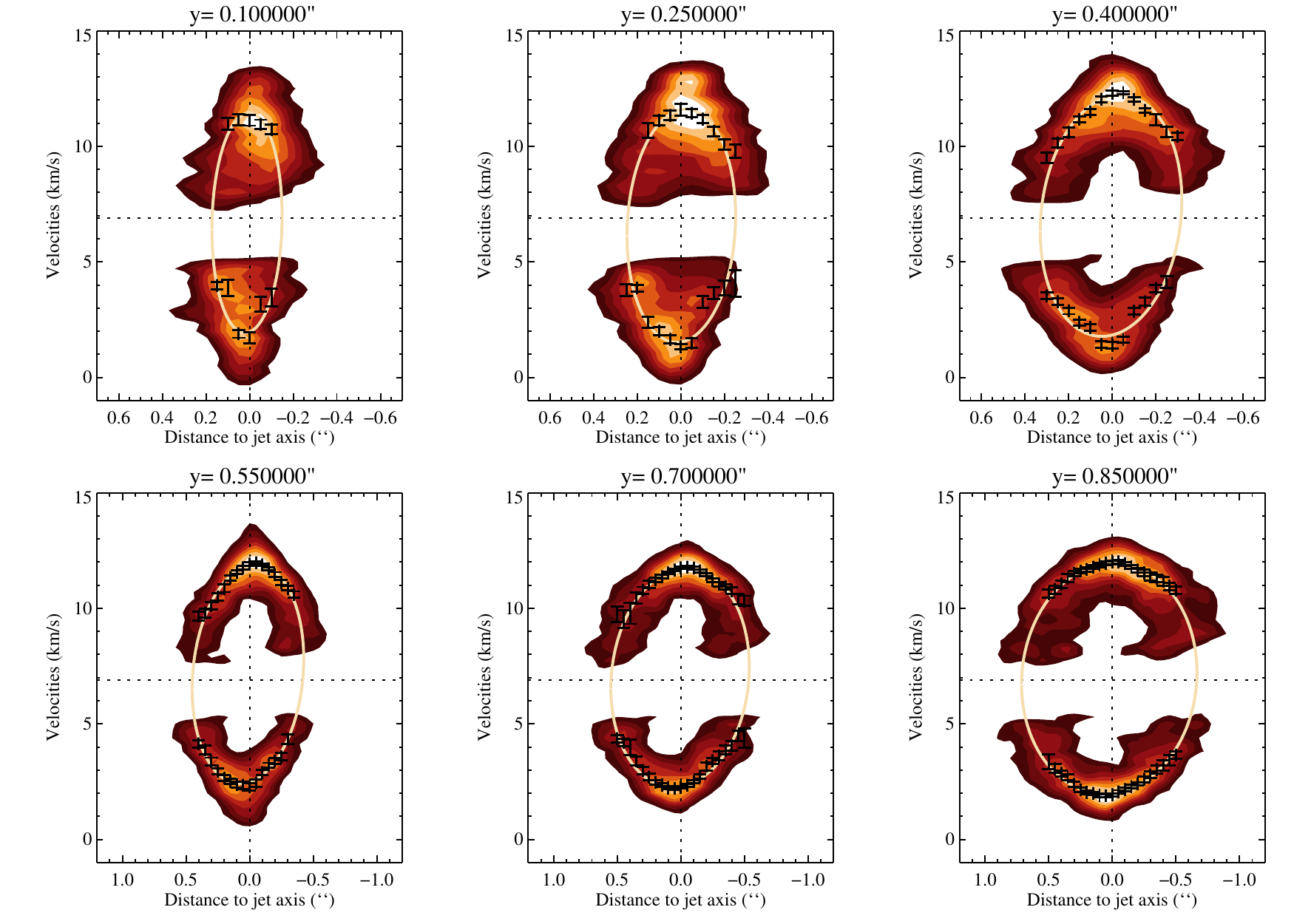}}}
\caption{Transverse pv diagrams fits from y=+0.1$^{\prime\prime}$ to y=+0.85$^{\prime\prime}$. The black crosses show the trace determined according to the method described in text. The white line shows the result of the fit of this trace by an ellipse. See text for more details.}
\label{f:extra-fit-ellipse1}
\end{figure*}

\begin{figure*}
\centerline{
\subfloat{\includegraphics[width=1\textwidth]{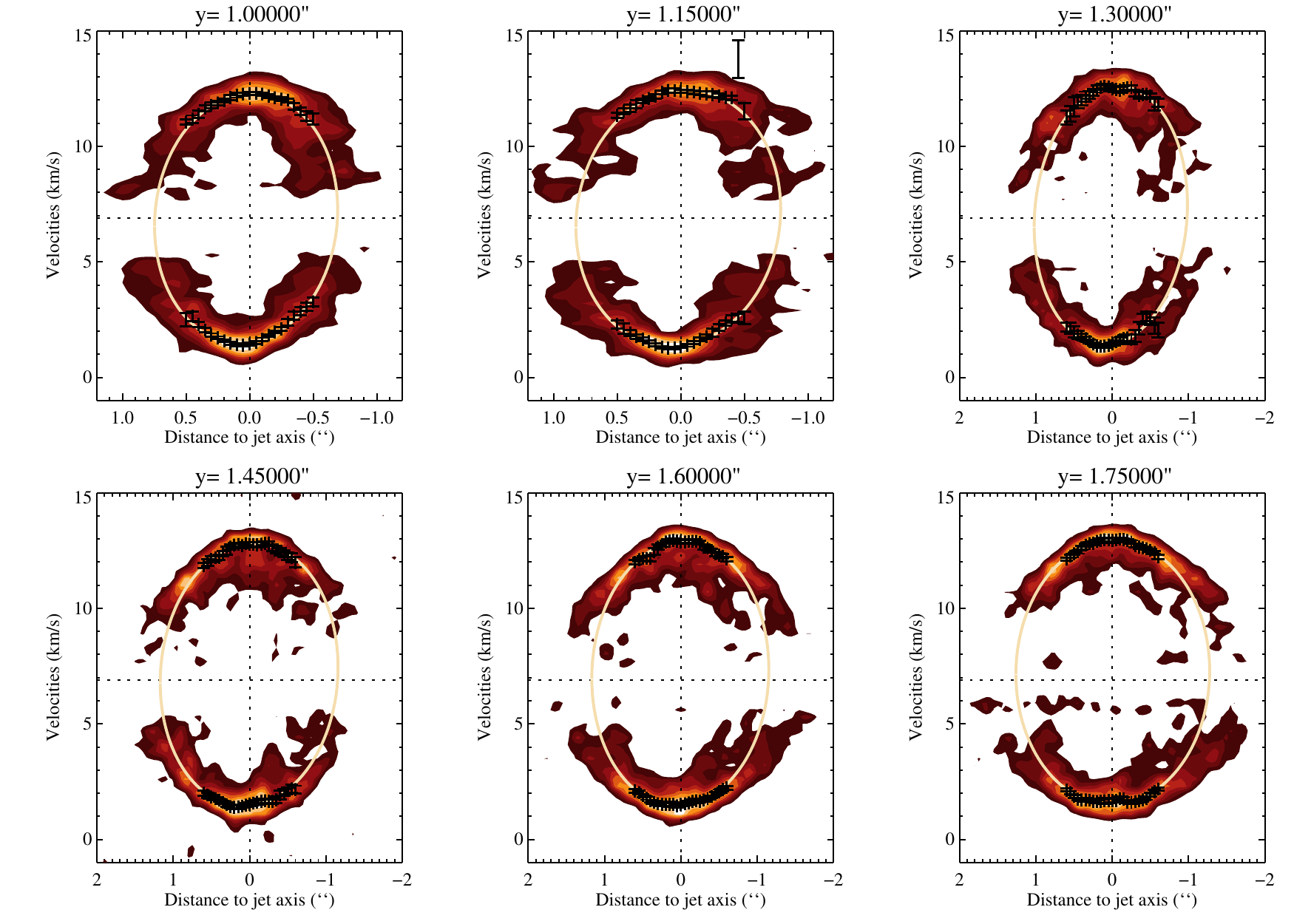}}}
\caption{Transverse pv diagrams fits from y=+1$^{\prime\prime}$ to y=+1.75$^{\prime\prime}$. The black crosses show the trace determined according to the method described in text. The white line shows the result of the fit of this trace by an ellipse. See text for more details.}
\label{f:extra-fit-ellipse2}
\end{figure*}

\section{Fitting the transverse position-velocity diagrams}

\label{ap:ringfit}

We give below the equations used in Section~4 to model the transverse pv diagrams. The procedure is very similar to that recently published in \citet{hirota17} except that we also allow for de-centring of the shell in order to retrieve possible wiggling signatures. We define (x,y) as the plane of the disk (see Fig.~\ref{fig:fitting}-left). The z-axis is assumed to be inclined at an angle $i$ with respect to the line of sight. In the case of HH 30, i $\simeq$ 90$^{\circ}$ so the plane of the sky is very close to the (x,z) plane. We assume that, at each altitude z above the plane of the disk, the flow can be modelled by a circular shell of radius $R$ with axisymmetric velocity components V$_z$, V$_r$, and V$_{\phi}$ in cylindrical coordinates. Due to wiggling, the centre of the shell can be offset in the x-axis by an amount $x_{\rm offset}$ (offsets in the y direction  will not be detectable). In the transverse pv diagram, constructed with the slit aligned along the x-axis (i.e. perpendicular to the jet), this shell of material projects onto an ellipse according to the following equations:

\begin{eqnarray*}
\rm
x-x_{\rm offset} & = & R \times \cos{\phi} \\
\rm V_{\rm los} - V_{0} & = & -\left( V_z \times \cos{i} + V_r \times \sin{i} \times \sin{\phi} + V_\phi \times \sin{i} \times \cos{\phi} \right),\\
\end{eqnarray*}

where $\phi$ is the azimuthal angle around the z-axis counted counterclockwise, V$_{\rm los}$ is the projected flow velocity along the line of sight (negative for approaching flow), and V$_{0}$ is the projected source velocity along the line of sight. 

By convention,V$_{\rm z}$ is positive for outward directed velocity component along the z-axis and V$_{\rm \phi}$ is positive for a rotation sense counterclockwise in the (x,y) plane, i.e. in the case of HH 30 for a rotation sense identical to the disk rotation sense measured in $^{13}$CO (see Fig. \ref{f:13co}).   

The relationships relating the five parameters of the ellipse (half axes a (major) and b (minor), centre coordinates r$_{cent}$, and V$_{cent}$, position angle PA) to the five parameters of the shell (R,x$_{\rm offset}$,V$_{\rm z}$,V$_{\rm r}$, V$_{\rm \phi}$) are given below:

\begin{eqnarray*}
x_{\rm offset} & = & r_{cent} \\
V{\rm_z} & = & -(V_{cent}-V_0)/\cos{i}\\
(V_{\rm r} \sin{i})^{2} & = & \left((\cos{PA})^2/a^2 + (\sin{PA})^2/b^2 \right)^{-1} \\
(V_{\phi} \sin{i})/R & = & 0.5 \times (V_{\rm r} \sin{i})^2 \times \sin{2 PA} \times (1/b^2 -1/a^2) \\
1/R^2 & = & \left( (\cos{PA})^2/b^2+(\sin{PA})^2/a^2 \right)-(V_{\rm \phi}/R)^2/V_{\rm r}^2. \\
\end{eqnarray*}

The position angle PA of the ellipse is measured from the $+V_{los}$ axis and counted positive towards the $ - x$ direction"


\end{appendix}

\end{document}